\documentclass[11pt]{article}
\usepackage[latin1]{inputenc}
\usepackage{amsmath}
\usepackage{amsfonts}
\usepackage{amssymb}
\usepackage{pstricks}
\usepackage{amsthm}
\usepackage{mathrsfs}
\usepackage{amssymb}
\usepackage{cancel}
\usepackage{slashed}
\usepackage{graphicx}
\usepackage[font=small,labelfont=bf]{caption}
\usepackage{float}
\usepackage{csquotes}
\usepackage[bottom]{footmisc} 
\usepackage{setspace}
\usepackage{bbding}
\usepackage{cite}
\usepackage{framed}
\usepackage{color}
\usepackage{wrapfig}
\definecolor{shadecolor}{RGB}{224,238,238}
\newcommand{\nn}{\nonumber}

\newcommand{\M}{\mathcal{M}}

\makeatletter

\@addtoreset{equation}{section}
\makeatother

\def\lsim{\;\raise0.3ex\hbox{$<$\kern-0.75em\raise-1.1ex\hbox{$\sim$}}\;}
\def\gsim{\;\raise0.3ex\hbox{$>$\kern-0.75em\raise-1.1ex\hbox{$\sim$}}\;}
\def\beq{\begin{equation}}   \def\eeq{\end{equation}}
\def\ba{\begin{array}}       \def\ea{\end{array}}
\def\bea{\begin{eqnarray}}   \def\eea{\end{eqnarray}}
\def\nn{\nonumber}
\def\nl{\newline}

\theoremstyle{definition} 
\date{\today}

\begin{document}

\begin{titlepage}
\begin{flushright}
LPT Orsay 15-97 \\
\end{flushright}


\begin{center}

\begin{doublespace}

\vspace{1cm}
{\Large\bf Discovery Prospects of a Light Scalar in the NMSSM} \\
\vspace{2cm}

{\bf{Ulrich Ellwanger$^{a,b}$ and Mat\'ias Rodr\'iguez-V\'azquez$^a$}}\\
\vspace{1cm}
{\it  $^a$ Laboratoire de Physique Th\'eorique, UMR 8627, CNRS, Universit\'e de Paris-Sud, Universit\'e
Paris-Saclay, 91405 Orsay, France\\
\it $^b$ School of Physics and Astronomy, University of Southampton,\\
\it Highfield, Southampton SO17 1BJ, UK}
\end{doublespace}
\end{center}
\vspace*{2cm}

\begin{abstract}
We study a region in the NMSSM parameter space in which the mass
of the SM-like Higgs boson is uplifted by $\sim 4-17$~GeV, allowing for
stop masses and $|A_t| \leq 1$~TeV alleviating the little fine tuning
problem of the MSSM. An uplift of the mass of the SM-like Higgs boson
is possible in two distinct regions in the NMSSM parameter space:
Either for large $\lambda$ and small $\tan\beta$ or, through singlet-doublet
mixing, for small $\lambda$ and large $\tan\beta$. For a
mostly singlet-like Higgs state $H_S$ with a mass
below $125$~GeV we investigate possible direct or
indirect search channels at the run~II of the LHC as function
of the NMSSM-specific uplift of the mass of the SM-like Higgs boson:
Direct production of $H_S$ in gluon fusion with $H_S$ decaying into
diphotons, modified reduced couplings of the SM-like Higgs state,
and the possible production of $H_S$ in $ggF \to A \to Z + H_S$.
We find that the region featuring singlet-doublet
mixing can be tested if searches at the LHC at 13~TeV for BSM Higgs
bosons in the mass range $88 - 102$~GeV decaying into diphotons
become sensitive to signal cross sections
$\sigma({gg\rightarrow H_S \rightarrow \gamma\gamma}) \sim 20$~fb, or
if measurements of the reduced coupling $\kappa_V(H_{SM})$ of the SM
Higgs boson to electroweak gauge boson exclude (or confirm) the region
$\kappa_V(H_{SM}) \lsim 0.93$.
\end{abstract}

\end{titlepage}
\newpage
\section{Introduction}
\label{sec:intro}

Since the discovery in 2012 of a SM-like Higgs scalar with a mass close to 125 GeV by 
the ATLAS \cite{Aad:2012tfa} and CMS \cite{Chatrchyan:2012xdj} collaborations,
its couplings to gauge bosons and fermions have been measured with an
unexpectedly high precision, see \cite{ATLAS-CMS} for a recent combination
of the measurements of ATLAS and CMS. These confirm essentially the
couplings expected from the Standard Model (SM).

Within supersymmetric extensions of the SM one finds an enlarged Higgs
sector featuring additional neutral CP-even, CP-odd and charged states.
It is relatively natural within the parameter space of supersymmetric
extensions of the SM to find a neutral CP-even Higgs state with couplings
to gauge bosons and fermions very close to the ones expected from the SM.
However, within the Minimal Supersymmetric extension of the SM (MSSM)
the mass of $\sim 125$~GeV of this SM-like Higgs state is not easy to
explain. At tree level the mass of the SM-like Higgs state is bounded
from above by $M_Z$, and accordingly large radiative corrections requiring large
scalar top (stop) masses and/or mass splittings well
above 1~TeV are needed in order to uplift the mass of the SM-like Higgs state
from $M_Z$ to $\sim 125$~GeV \cite{Hall:2011aa,Baer:2011ab,Heinemeyer:2011aa,
Arbey:2011ab,Draper:2011aa, Carena:2011aa,Buchmueller:2011ab,Arvanitaki:2011ck}.

But heavy stop masses/mass splittings
lead to large radiative corrections to a soft Susy breaking Higgs mass
term, which has to be tuned against the $\mu$ parameter if much larger
than $M_Z$ (see \cite{Papucci:2011wy} and refs. therein). Accordingly the Higgs mass of
about 125~GeV aggravates a little finetuning problem within the MSSM,
pointed out already in the context of LEP bounds on the Higgs mass in
\cite{Chankowski:1997zh,Barbieri:1998uv,Kane:1998im,Giusti:1998gz}.

It is known that the upper tree level bound of $M_Z$ on the mass of the
SM-like Higgs boson does not hold in the Next-to-Minimal Supersymmetric
extension of the SM (NMSSM, see \cite{Maniatis:2009re,Ellwanger:2009dp}
for reviews). Accordingly the NMSSM can alleviate the little finetuning
problem of the MSSM \cite{Arvanitaki:2011ck,Papucci:2011wy,BasteroGil:2000bw,
Dermisek:2005ar,Dermisek:2005gg,
Dermisek:2007yt,Delgado:2010uj,Ellwanger:2011mu,Ross:2011xv,Kang:2012sy,Cao:2012yn,
Perelstein:2012qg}. 
It shares the benefits of supersymmetric extensions of the Standard
Model (SM) with the MSSM: The hierarchy problem can be strongly reduced,
the presence of dark matter can be explained, and the running gauge
couplings are automatically consistent with a Grand Unified Theory (GUT).

In the NMSSM, an additional gauge singlet superfield $\widehat{S}$ couples
with a dimensionless coupling $\lambda$ to the
two SU(2) doublet superfields $\widehat{H}_u$ and $\widehat{H}_d$ of the MSSM.
A vacuum expectation value of the scalar component of $\widehat{S}$
generates dynamically a $\mu$ parameter of the order of the Susy
breaking scale, solving the $\mu$-problem of the MSSM~\cite{Kim:1983dt}.
The NMSSM spectrum contains three neutral CP-even Higgs scalars.
Typically, one of them is mostly SM-like (denoted by $H_{SM}$ in the
following), one has the properties of the (heavy) MSSM-like state $H$,
and a third state $H_S$ is mostly singlet-like. These states are mixtures
of the weak eigenstates (the scalar components of $\widehat{H}_u$, $\widehat{H}_u$
and $\widehat{S}$). Past and present searches for Higgs bosons 
at LEP, the Tevatron and the LHC do not exclude masses of $H_S$
below 125~GeV. 

In fact, in the NMSSM two distinct mechanisms can
lead to additional tree level contributions to the mass of the
SM-like state $H_{SM}$:

a) If $\lambda$ is large enough ($\lambda^2 > (g_1^2+g_2^2)/2$, where
$g_1$ and $g_2$ are the electroweak gauge couplings) and $\tan\beta$ is
small enough ($\tan\beta \lesssim 6$), the additional quartic coupling
$\sim \lambda^2 H_{SM}^4$ in the scalar potential lifts its mass above $M_Z$. However,
$\lambda \gtrsim 1$ (so-called $\lambda$Susy 
\cite{Barbieri:2006bg,Barbieri:2013hxa}) would be required in order to
push the tree level mass from $M_Z$ to 125~GeV in which case $\lambda$ runs into
a Landau singularity well below the GUT scale. In order to avoid this
we confine ourselves subsequently to $\lambda \lesssim 0.75$.

b) If the mostly singlet-like state $H_S$ has a mass below 125~GeV,
mixing between $H_S$ and $H_{SM}$ (more precisely, among the weak
eigenstates) leads to an increase of the mass of the latter.
The impact of such mixings on the Higgs spectrum of the NMSSM has been
known for a while \cite{Delgado:2010uj,Ellwanger:2011mu,Barger:2006dh,
Ellwanger:2006rm,Dermisek:2007ah,
Barbieri:2007tu}, but became particularly
interesting once the mass of $\sim 125$~GeV of the mostly SM-like state had been measured
\cite{Hall:2011aa,Kang:2012sy,Cao:2012fz,Jeong:2012ma,Agashe:2012zq,
Choi:2012he,Kowalska:2012gs,King:2012tr,Kang:2013rj,Christensen:2013dra,Cheng:2013fma,
Badziak:2013bda,Beskidt:2013gia,Choi:2013lda,Cao:2013gba,Cacciapaglia:2013ora,
Englert:2014uua,Ellwanger:2014dfa,Badziak:2014aca,Jeong:2014xaa,King:2014xwa,Allanach:2015cia,
Buttazzo:2015bka,Guchait:2015owa,Domingo:2015eea}. The mass shift of up
to $\sim 8$~GeV occurs now mostly for large $\tan\beta$ and smaller
$\lambda \approx 0.04-0.1$, the latter in order to avoid constraints from
LEP on a Higgs-like state with a mass below $\sim 114$~GeV \cite{Schael:2006cr}.
(The increase of the mass of the SM-like state $H_{SM}$ through mixing implies a decrease of the
lighter singlet-like state $H_S$.) Hence the corresponding region in
parameter space is clearly distinct from the one where the quartic
SM-Higgs self coupling is enhanced.

In the present paper we consider both possibilities, but confine ourselves
to the case where the mass of the mostly singlet-like state $H_S$ is below 125~GeV:
This situation is preferred also in the large $\lambda$--small $\tan\beta$
regime, since singlet-doublet mixing would always imply a decrease of
the mass of the SM-like state if the singlet-like state is heavier, and
mixing is hard to avoid if $\lambda$ is large (unless $H_S$ is very heavy
and/or the corresponding off-diagonal element in the mass matrix happens
to be small).
On the other hand a mass of the mostly singlet-like state $H_S$ below
$\sim 60$~GeV would lead to dominant decays of $H_{SM}$ into pairs of
$H_S$ unless $\lambda$ (and hence the mixing angle) is very small; also
the LEP constraints are quite strong for this mass range \cite{Schael:2006cr}.
We found that a sizeable positive mass shift for the SM-like state is unlikely
here.

It is known that singlet-doublet mixing has two distinct phenomenological
consequences:

a) The mostly singlet-like state inherits couplings to SM gauge bosons
and fermions from the SM-like state proportional to the (sinus of the) mixing
angle. This leads to non-vanishing production cross sections for $H_S$,
and its potential discovery at the LHC.

b) Simultaneously, the couplings of $H_{SM}$ to gauge bosons
and fermions get reduced. The uncertainties of the measured couplings of $H_{SM}$ at
the run~I of the LHC \cite{ATLAS-CMS} are expected to decrease further
after measurements at the run~II \cite{ATL-PHYS-PUB-2013-014,CMS:2013xfa}.

It is the purpose of the present paper to study in how far the combination
of both sources of future information can constrain the presence --
or lead to a discovery -- of a light singlet-like Higgs boson in the NMSSM,
as function of the NMSSM specific mass shift of the SM-like state.
We also indicate the possible production of $H_S$ in decays of heavier
MSSM-like $H/A$ states.
To start with, we have to collect the available constraints on this
scenario from LEP and measurements at the run~I of the LHC.

First, bounds on couplings
to the $Z$~boson times the branching fraction of an additional light Higgs
boson into $b\bar{b}$ and gluons originate from LEP \cite{Schael:2006cr}.

Second, limits originate
from direct searches for extra (lighter) Higgs states in the diphoton channel
by ATLAS \cite{Aad:2014ioa} and CMS \cite{CMS-HIG-14-037}: despite the
relatively small diphoton branching fraction this final state
is the most promising one to search for, in particular in view of
the possibility that the diphoton branching fraction of $H_S$ can be
considerably larger than the one of a SM-like Higgs boson of corresponding
mass \cite{Christensen:2013dra,Badziak:2013bda,King:2014xwa,Guchait:2015owa,
Domingo:2015eea,
Moretti:2006sv,Ellwanger:2010nf,Cao:2011pg,Benbrik:2012rm,Heng:2012at,
Jia-Wei:2013eea}.

Third, limits originate from the potential reduction of couplings of
$H_{SM}$ to SM gauge bosons and fermions through mixing with a gauge
singlet. The corresponding measurements of production and decay mode
dependent signal strengths of ATLAS and CMS have recently
been combined by the collaborations in \cite{ATLAS-CMS}. Global fits to
the couplings (or the coupling modifiers) require, in principle,
likelihood grids including information on deviations from Gaussianity
and correlations among uncertainties in particular for identical
final states from different production modes. Moreover such
global fits depend crucially on the assumptions on the underlying model
like custodial symmetry (identical modifications of couplings to $W$ and
$Z$ bosons), correlated modifications of couplings to $b$~quarks and
$\tau$~leptons like in specific Higgs doublet models, and possible
additional contributions to loop~induced couplings to gluons and
photons. 

The latest global fits including assumptions corresponding
to the NMSSM (custodial symmetry, correlated modifications of couplings
to $b$~quarks and $\tau$~leptons, possible additional contributions
notably to the loop~induced coupling to photons) have been performed
in \cite{Bernon:2014vta}. We have checked that their combined signal
strengths are very close to the ones in \cite{ATLAS-CMS} and use,
for the scan of the NMSSM parameter space (see below), their
95\%~CL on signal strengths of $H_{SM}$ (verifying only subsequently
the bounds from \cite{ATLAS-CMS}). Electroweak
precision data (the $W$~boson mass) do not constrain the parameter
space of the NMSSM with a light $H_S$ \cite{Stal:2015zca}.
Overall, in the NMSSM the experimental constraints on the 
$H_{SM}-H_S$ mixing angle (for $M_{H_S}$ below 125~GeV) are similar to
the ones obtained from studies within simple singlet-extensions of the
non-supersymmetric SM \cite{Buttazzo:2015bka,Robens:2015gla,Falkowski:2015iwa,
Gorbahn:2015gxa,Godunov:2015nea}.

As a next step we study in how far future measurements of diphoton signal
rates (via ggF) of $H_S$ at 13~TeV  are sensitive to the NMSSM specific
mass shift of the SM-like state. Likewise, the dependence of the
couplings of $H_{SM}$ (and hence of the $H_{SM}-H_S$ mixing angle) 
on the NMSSM specific mass shift of the SM-like state is
analysed. The results clarify in how far the NMSSM specific mass shift
can be tested in the future, and which of the different measurements
are potentially more sensitive. Studies of possible $H_S$ diphoton signal
rates (after the discovery of the $H_{SM}$ state) in the NMSSM have
been performed earlier in \cite{Christensen:2013dra,Badziak:2013bda,
Jia-Wei:2013eea,Guchait:2015owa,Domingo:2015eea}
(see also \cite{Buttazzo:2015bka}), and correlations
with mass shifts (from $H_{SM}-H_S$ mixing only)  have been presented in \cite{Badziak:2014aca}.
In the present paper we extend the studies of such correlations
including the large $\lambda$-small $\tan\beta$ regime,
include constraints
from ATLAS \cite{Aad:2014ioa} and CMS \cite{CMS-HIG-14-037}
from direct searches for lighter Higgs states in the diphoton channel,
and obtain possible $H_S$ diphoton signal rates which partially
deviate from (are larger than) the ones obtained earlier.

In the next section we recall the properties of the Higgs sector of the
NMSSM relevant for the present study, and define a NMSSM specific mass
shift $\Delta_\mathrm{NMSSM}$ of the SM-like Higgs state. In section~3
we describe the scans over the parameter space. In section~4 we present
the results of the scans as function of $\Delta_\mathrm{NMSSM}$:
$H_S$ diphoton signal rates at 8 and 13~TeV c.m. energy,
modifications of the couplings of  $H_{SM}$, and correlations among them.
We discuss and
compare prospects for tests of the scenarios under study, including the possible
production of $H_S$ in decays of heavy MSSM-like $H/A$ states. Finally
we conclude in section~5.

\section{The neutral Higgs sector of the NMSSM}
\label{sec:thehiggssector}
In this paper we consider the CP-conserving $\mathbb{Z}_3$-invariant NMSSM.
The superpotential of the NMSSM Higgs sector reads
\beq
W_{\text{Higgs}}=\lambda\hat{S}\hat{H}_u\cdot \hat{H}_d+\frac{\kappa^3}{3}\hat{S}^3
\label{2.1e}
\eeq
where $\hat{S}$ is the chiral singlet superfield. Once the real
component of the superfield $\hat S$ develops a vacuum expectation value (vev) $s$,
the first term in the
superpotential generates an effective $\mu$ term
\beq
\mu= \lambda s\; .
\eeq
The soft Higgs-dependent SUSY breaking terms are
\beq
\mathcal{L}_{\text{Soft}}=-m^2_{H_u}|H_u|^2 - m^2_{H_d}|H_d|^2-m_S^2|S|^2
-\left(\lambda A_\lambda H_u\cdot H_d S
+ \frac{1}{3}\kappa A_\kappa S^3 +\text{h.c}\right)\; .
\eeq
Then, from the SUSY gauge interactions, the $F$ and soft SUSY breaking terms
one obtains the Higgs potential 
\begin{eqnarray}
V &=& |\lambda \left(H^+_u H_d^- - H^0_u H^0_d\right) + \kappa  S^2|^2 \nn\\
&&+ \left( m^2_{H_u} +|\mu+\lambda S|^2\right)\left(|H^0_u|^2+ |H^+_u|^2\right)^2+ \left( m^2_{H_d}
+|\mu+\lambda S|^2\right)\left(|H^0_d|^2+ |H^-_d|^2\right)^2 \nn\\
&&\frac{g_1^2+g_2^2}{8}\left( |H^0_u|^2 + |H^+_u|^2 - |H^0_d|^2- |H^-_d|^2\right)^2 + 
\frac{g_2^2}{2}|H_u^+H^{0*}_d+H^0_u H^{-*}_d|^2 \nn\\
&& + m_S^2 |S|^2 + \left( \lambda A_\lambda(H_u^+H^-_d-H^0_u H^0_d)S +
\frac{1}{3}\kappa A_\kappa S^3 + \text{h.c.}\right).
\end{eqnarray}
After expanding around the vacuum expectation values $v_u$, $v_d$ and $s$
(which can be taken to be real and positive),
 the Higgs fields are given by
\bea
H_u=\left(\begin{array}{c}
H^+_u \\
H^0_u= v_u + \frac{1}{\sqrt{2}}(H^0_{u,r} + i H^0_{u,i})
\end{array}\right) ,\qquad
&& H_d=\left(\begin{array}{c}
H^0_d= v_d + \frac{1}{\sqrt{2}}(H^0_{d,r} + i H^0_{d,i}) \\
H^-_d 
\end{array}\right)\; ,
\nn\\
S=s+\frac{1}{\sqrt{2}}(S_r +i S_i)\; .&&
\eea
Once the soft Higgs masses are expressed in terms of $M_Z$, $\tan\beta$ and $s$ using the minimization
equations of the potential, the Higgs sector of
the NMSSM at tree level is described by six parameters
\beq
\lambda, \quad\kappa, \quad\tan\beta,\quad \mu=\lambda s, \quad A_\lambda\quad \text{and}\quad A_\kappa\; .
\label{2.6e}
\eeq
Defining $v^2=2M_Z^2/(g_1^2+g_2^2)\sim (174\ \mathrm{GeV})^2$, the $3\times 3$
CP-even mass matrix in the basis
$\left(H_{d,r}, H_{u,r}, S_r\right)$ reads:
\begin{eqnarray}
 \M^2_{S,11} &=&	M_Z^2  \cos^2\beta + \mu (A_\lambda+\kappa s)\tan\beta\;  ,\nn\\
 \M^2_{S,12}&=&(\lambda v^2-\frac{M_Z^2}{2})\sin 2\beta-\mu (A_\lambda+\kappa s)\; ,\nn\\
 \M^2_{S,13}&=& \lambda v \left(2\mu \cos\beta - (A_\lambda+2\kappa s )\sin\beta \right))\; ,\nn\\
 \M^2_{S,22} &=&  M_Z^2 \sin^2\beta +  \mu (A_\lambda+\kappa s)\cot\beta + \Delta_{\text{rad}}\; ,\nn\\
 \M^2_{S,23} &=&  \lambda v \left(2\mu \sin\beta - (A_\lambda+2\kappa s)\cos\beta\right))\; ,\nn\\
 \M^2_{S,33} &=& \lambda A_\lambda \frac{v^2}{2s}\sin 2\beta  + \kappa s(A_\kappa+4\kappa s)\; .
\end{eqnarray}
Here $\Delta_{\text{rad}}$ denotes the dominant radiative corrections due to top/stop loops,
\beq
 \Delta_{\text{rad}}=\frac{3 m_t^4}{4\pi^2 v^2}\left(\ln\left(\frac{m_T^2}{m_t^2}\right)+\frac{X_t^2}{m_T^2}
 \left(1-\frac{X_t^2}{12 m_T^2}\right)\right)
\eeq
where $m_T$ is the geometrical average of the soft SUSY breaking stop masses, and $X_t = A_t -
\mu/\tan\beta$ with $A_t$ the soft SUSY breaking stop trilinear coupling.

It is convenient to rotate $\M^2_S$ by an angle $\beta$ in the doublet sector sector
into $\M'^2_S$ in the basis $H_{SM}',H',S'$ (with $S' \equiv S_r$):
\beq
\M'^2_{S} = R(\beta) \M^2_S R^{\cal T}(\beta)\; , \qquad
R(\beta) = \left(\ba{ccc} \cos\beta & \sin\beta & 0 \\ 
\sin\beta & -\cos\beta & 0\\
0 & 0 & 1 \ea\right)\; .
\eeq
 Such a basis (also known as Higgs basis)
has the advantage that only the component $H_{SM}'$ of the Higgs doublets acquires
a vev $v$ and that, for realistic parameters, it is nearly diagonal: $H_{SM}'$ has SM-like
couplings to fermions and electroweak gauge bosons,
the  heavy doublet field $H'$ is the CP-even partner of the MSSM-like CP-odd state $A$,
while $S'$ remains a pure singlet. The mass matrix
$\M'^2_S$ in the basis $(H_{SM}',H',S')$ has the elements
\begin{eqnarray}
\M'^2_{S,11}&=& M^2_Z\cos^2 2\beta + \lambda^2 v^2 \sin^2 2\beta + \sin^2\beta \Delta_{\text{rad}}\; ,
\nn \\
\M'^2_{S,12}&=& \sin 2\beta\left( \cos 2\beta \left(M_Z^2-\lambda^2 v^2\right) 
- \frac{1}{2}\Delta_{\text{rad}}\right)\; ,\nn \\
\M'^2_{S,13}&=& \lambda v \left(2\mu -(A_{\lambda} + 2\nu) \sin 2\beta\right)\; ,\nn\\
\M'^2_{S,22}&=& M_A^2+\left(M_Z^2-\lambda^2 v^2 \right)\sin^2 2\beta + \cos^2\beta \Delta_{\text{rad}}
\; ,\nn\\
\M'^2_{S,23}&=& \lambda v(A_{\lambda} + 2\nu) \cos 2\beta\; ,\nn \\
\M'^2_{S,33}&=& \lambda A_\lambda\frac{v^2}{2s} \sin 2\beta + \nu\left(A_\kappa + 4\nu\right)\; ,
\label{2.10e}
\end{eqnarray}
where we have defined $\nu = \kappa s$ and
\beq
M_A^2=\frac{2\mu}{\sin 2\beta}(A_\lambda+\nu)\; ,
\eeq
the mass squared of the MSSM-like CP-odd state $A$. ($A$ mixes, in principle,
with a mostly singlet-like state $A_S$. We will comment on the mass range of
the CP-odd states in section~4.3.)

After an additional final diagonalisation
the eigenstates will be denoted as
\begin{itemize}
\item $H_{SM}$ \text{(dominantly\ SM-like)}
\item $H_S$ \text{(dominantly\ singlet-like)}
\ \text{and}
\item $H$ \text{(dominantly the MSSM-like heavy scalar).}
\end{itemize}

By this final diagonalisation the state $H_S$ picks up couplings to
electroweak gauge bosons (vector bosons)
proportional to the $H_{SM}'-S'$ mixing angle. Defining by $\kappa_V$ the ratio of couplings
of a Higgs state to vector bosons relative to the corresponding coupling of the
SM-like Higgs boson, one has
\beq
\kappa_V^2(H_{SM})+\kappa_V^2(H_S)+\kappa_V^2(H)=1\; .
\eeq
$H_{SM}'-S'$ mixing will necessarily generate $\kappa_V^2(H_{S})\neq 0$
and hence reduce $\kappa_V^2(H_{SM})$, which is already
\cite{ATLAS-CMS}
and will be even more constrained by Higgs coupling measurements at the LHC.
Similarly, the state $H_S$ picks up couplings to fermions by both 
$H_{SM}'-S'$ and $H'-S'$ mixing, leading to non-vanishing values for
$\kappa_U(H_S)$ (the reduced coupling of $H_S$ to up-type quarks) and
$\kappa_D(H_S)$ (the reduced coupling of $H_S$ to down-type quarks). Then
loop diagrams generate non-vanishing values for
$\kappa_{gg}(H_S)$ (the reduced coupling of $H_S$ to gluons) and
$\kappa_{\gamma\gamma}(H_S)$ (the reduced coupling of $H_S$ to diphotons).
It is important to note that the coupling of $H_S$ to down-type quarks
can suffer from cancellations among the contributions from
$H_{SM}'-S'$ and $H'-S'$ mixing, respectively \cite{Ellwanger:2010nf}. This
can result in a reduced branching fraction $BR(H_S \to b\bar{b})$. Since
this decay constitutes the dominant contribution to the total width of
$H_S$, its reduction implies enhanced branching fractions into other
final states like $\gamma\gamma$. It is thus not astonishing that the
$BR(H_S \to \gamma\gamma)$ can be larger than the one of a SM-Higgs boson of
corresponding mass, leading to $\kappa_{\gamma\gamma}(H_S) > 1$.

The diagonal term in (\ref{2.10e}) associated with the mass of the mostly SM Higgs is
\begin{equation}
\M'^2_{S,11}= M^2_Z\cos^2 2\beta + \lambda^2 v^2 \sin^2 2\beta + \sin^2\beta \Delta_{\text{rad}}
\label{eq:m11}
\end{equation}
where the first term is the tree level upper bound for the Higgs mass in the MSSM. Due to the wide mass
gap between $M_Z$ and $\sim 125$~GeV it is necessary to consider mechanisms able to uplift the Higgs mass
from its MSSM-like tree level value. In the MSSM this may be achieved by sizeable
radiative corrections $\Delta_{\text{rad}}$ which require large ($\gg 1$~TeV) values for at
least one soft SUSY breaking stop mass term and/or $A_t$. Such soft SUSY breaking terms
generate, via loop effects, a soft SUSY breaking Higgs mass term $m_{H_u}^2$ ($< 0$) of the same order.
On the other hand, combining the (tree level) minimisation equations of the potential for the vevs
$v_u$ and $v_d$, one obtains
\beq
M_Z^2=\frac{2(m^2_{H_d}- m^2_{H_u}\tan^2\beta)}{\tan^2\beta -1}-2\mu^2\; .
\eeq
In the absence of fine tuning, no large cancellations between the terms on the right hand side should occur.
Hence large radiative corrections $\Delta_{\text{rad}}$ generate a so-called
``little fine tuning problem'' in the MSSM
\cite{Papucci:2011wy,Chankowski:1997zh,Barbieri:1998uv,Kane:1998im,Giusti:1998gz}.
Moreover, the (effective) $\mu$ parameter should not be much larger than $M_Z$.

The second term in (\ref{eq:m11}) is the well known NMSSM-specific contribution to the
SM-like Higgs mass \cite{Maniatis:2009re,Ellwanger:2009dp}, which is numerically
relevant for $\tan\beta \lsim 6$ and large
$\lambda$. Avoiding a Landau singularity below the GUT scale requires $\lambda \lsim 0.75$,
limiting the possible uplift of the mass of the SM-like Higgs state to $\lsim 17$~GeV.

A third possibility to uplift the mass of the SM-like Higgs state has recently been
studied in some detail in \cite{Badziak:2013bda,Badziak:2014aca}: If 
the diagonal term $\M'^2_{S,33}$ in (\ref{2.10e}) associated with the mass of
the singlet-like Higgs state $S'$ is smaller than $\M'^2_{S,11}$,
$H_{SM}' - S'$ mixing induced by the term $\M'^2_{S,13}$ in (\ref{2.10e}) shifts
upwards the mass of the SM-like Higgs state $H_{SM}$. The dominant contribution
to $\M'^2_{S,13}$ originates from the first term $2\lambda v \mu$, which
gets reduced by the second term $-\lambda v (A_{\lambda} + 2\nu) \sin 2\beta$. This
reduction becomes small for moderate to large values of $\tan\beta$
\cite{Badziak:2013bda,Badziak:2014aca}. On the other hand,  $H_{SM}' - S'$ mixing
induces couplings of the lighter eigenstate $H_S$ to electroweak gauge bosons,
$b\bar{b}$ and gluons (through top quark loops). Such couplings of a state
with a mass below 114~GeV are constrained by LEP \cite{Schael:2006cr}.
This limits the region of $\lambda$ for a sizeable uplift the mass of the
SM-like Higgs state to $\lambda \sim 0.04 ... 0.1$,
and the possible uplift the mass of the SM-like Higgs state to $\lsim 8$~GeV
 \cite{Badziak:2013bda,Badziak:2014aca}.

Subsequently we intend to quantify the NMSSM-specific uplifts of the
the mass of the SM-like Higgs state. To this end we define a mass shift
$\Delta_\mathrm{NMSSM}$ of the mostly SM-like Higgs state due to the NMSSM
specific effects, from the second term in (\ref{eq:m11}) and/or from
$H_{SM}' - S'$ mixing. Contributions from $H_{SM}' - S'$ mixing are easy to identify;
it suffices to compare the second eigenvalue of ${\cal M}_S^2$ (corresponding to $M^2_{H_{SM}}$)
to the case where $\lambda,\ \kappa \to 0$ (keeping $\mu$ fixed, which requires to keep
the ratio $\kappa/\lambda$ fixed). Such a definition of
$\Delta_\mathrm{NMSSM}$ has already been employed in
\cite{Badziak:2013bda,Badziak:2014aca}. In addition we want to keep track of the NMSSM contribution 
from the second term in (\ref{eq:m11}) relative
to the MSSM, which is relevant for small $\tan\beta$ only. But keeping small $\tan\beta$
would reduce the MSSM-like tree level value $M_Z^2 \cos^2 2\beta$, and it would not be
``fair'' to compare the NMSSM to the MSSM for low
values of $\tan\beta$. Hence we evaluate the contribution to $\Delta_\mathrm{NMSSM}$ in the
large $\lambda$-low $\tan\beta$ regime of the NMSSM by comparing to the MSSM
($\lambda,\ \kappa \to 0$ as before)
with a large value of $\tan\beta = 40$. (The SM-like Higgs mass in the MSSM
is practically independent of $\tan\beta$ for $\tan\beta > 40$.)
Therefore, for a given set of parameters in (\ref{2.6e}), 
\bea
\Delta_\mathrm{NMSSM} =M_{H_{SM}}- \max_{\tan\beta} M_{H_{SM}}
\bigg|_{\lambda,\kappa\rightarrow 0}
\simeq  M_{H_{SM}}-  M_{H_{SM}} \bigg|_{
\lambda,\kappa\rightarrow 0,\ \tan\beta=40}\; .
\eea
Clearly, larger values of $\Delta_\mathrm{NMSSM}$ require smaller radiative corrections
$\Delta_{\text{rad}}$ to $\M'^2_{S,11}$ and alleviate correspondingly the
little hierarchy problem. Accordingly $\Delta_\mathrm{NMSSM}$ can be
interpreted as an approximate measure of naturalness.

It is the aim of the present paper to study in how far such natural regions
in the parameter space of the NMSSM can be tested in the future, as function
of $\Delta_\mathrm{NMSSM}$ and the mechanism for an NMSSM-specific
uplift of the mass of the SM-like Higgs state.
Since $H_{SM}' - S'$ mixing has a negative effect on $\Delta_\mathrm{NMSSM}$
for $M_{H_S} > 125$~GeV (also if $\Delta_\mathrm{NMSSM}$ originates
mainly from the second
term in (\ref{eq:m11})) we will concentrate on $M_{H_S} < 125$~GeV.
Then, present constraints and future discoveries/constraints can originate
from
\begin{itemize}
\item direct searches for $H_S$ in the diphoton final state, which had
been carried out by ATLAS for
$65\ \mathrm{GeV} < M_{H_S}$ \cite{Aad:2014ioa} and by CMS for
$80\ \mathrm{GeV} < M_{H_S} < 115\ \mathrm{GeV}$ \cite{CMS-HIG-14-037}.
\item measurements of the reduced signal rates/couplings (with respect to
the SM) of $H_{SM}$. In the case of $H_{SM}' - S'$ mixing, these signal rates/couplings
diminish proportional to the mixing angle.
\item possible production of $H_S$ in decays
of the MSSM-like states $H/A$.
\end{itemize}
Comparing the corresponding sensitivities allows to verify under which
conditions natural NMSSM scenarios with $M_{H_S} < 125\ \mathrm{GeV}$
can be tested at future runs at the LHC, depending on
the mechanism for the NMSSM-specific uplift of the mass of the SM-like Higgs state.
To this end we have scanned the parameter space of the NMSSM as described in
the next section.

\section{Numerical analysis}
\label{sec:numerical}
We have performed these calculations with the public code
\texttt{NMSSMTools\_4.4.0}
 \cite{Ellwanger:2004xm,Ellwanger:2005dv} including
up to two-loop radiative corrections to the Higgs mass matrices as
obtained in \cite{Degrassi:2009yq}. All phenomenological constraints,
including the absence of Landau singularities below the GUT scale and,
notably, constraints from Higgs searches in various channels at LEP
are applied as in NMSSMTools (except for $(g-2)_\mu$).

The NMSSM specific parameters in Eq.~(\ref{eq:m11}) are varied  in the ranges 
\bea
0.001 \leq \lambda < 0.75, &0.001 \leq \kappa \leq 0.75,& 1 \leq \tan\beta \leq 50, \nn\\
0 \leq A_\lambda \leq 2.5~\mathrm{TeV}, &-1~\mathrm{TeV} \leq A_\kappa \leq 0,&
100~\mathrm{GeV} \leq \mu \leq 250~\mathrm{GeV}\; ;
\label{2.8e}
\eea
we found that wider ranges of the trilinear couplings $A_\lambda$, $A_\kappa$ and $\mu$
(including negative values of $\kappa$ and/or $\mu$) have
practically no impact on our results. The soft SUSY breaking squark masses of the
third generation $M_{U_3},\ M_{D_3},\ M_{Q_3}$ and the stop mixing parameter $A_t$
are confined to ranges below 1~TeV in order to avoid too large fine tuning: 
\beq\label{eqn:stopmassrange}
700~\mathrm{GeV} \leq M_{U_3}=M_{D_3}=M_{Q_3} \leq 1~\mathrm{TeV}, \qquad
-1 \text{ TeV} \leq A_t \leq 1~\mathrm{TeV}\; .
\eeq
(For $|A_t| \leq 1~\mathrm{TeV}$, third generation squark masses
below $\sim 700$~GeV do not allow to
reach  $125.1 \pm 3$~GeV for $M_{H_{SM}}$ even in the NMSSM.) The lightest physical
stop mass $m_{\tilde{t}_1}$ satisfies $ m_{\tilde{t}_1} \gsim 480$~GeV.

The soft SUSY breaking mass terms and trilinear couplings for the sleptons have been
set to  $500$ GeV and $550$ GeV respectively, whereas for the squarks of first two generations the masses
are set to 2~TeV. The gluino mass is chosen as $M_3=1.6$~TeV, and the
other soft SUSY breaking gaugino masses such that they satisfy approximately universal relations at the GUT
scale, i.e. $M_2=2M_1=M_3/3$. (All these parameters have practically no impact
on our results.)

For each point in the parameter space satisfying the phenomenological constraints, including a SM-like Higgs
state with a mass of $125.1 \pm 3$~GeV (allowing for theoretical errors)
and couplings of $H_{SM}$ to gauge bosons and
fermions in the 95\%~CL ranges given in \cite{ATLAS-CMS,Bernon:2014vta}, we further require  $M_{H_s} < M_{H_{SM}}$.  
Then we compute for each point  $\Delta_\mathrm{NMSSM}$ according to the procedure described above, and
various observables like reduced couplings and signal rates for the
relevant Higgs states shown in the next section. 

%
\section{Results}
\label{sec:results}
Due to the limited range \eqref{eqn:stopmassrange} for the soft SUSY breaking squark masses of the
third generation and the stop mixing parameter, all viable points
need a non-vanishing value of $\Delta_\mathrm{NMSSM}$ in the range $4~\text{GeV}\lesssim
\Delta_\mathrm{NMSSM} \lesssim 17~\text{GeV}$ in order reach a SM-like Higgs mass of $125.1 \pm 3$~GeV.
Hence this range for the soft SUSY breaking squark masses of the
third generation and the stop mixing parameter, motivated by alleviating the little hierarchy problem,
is not viable in the MSSM.

Turning to the possible mechanisms for an uplift of the mass of the SM-like Higgs state, it follows
from the discussion in section \ref{sec:thehiggssector} that these take place in different
regions of $\lambda$ and $\tan\beta$: contributions to $\Delta_\mathrm{NMSSM}$ up to $\sim 17$~GeV
from the second term
in (\ref{eq:m11}) (limited by the absence of a Landau singularity of $\lambda$ below the GUT
scale) are possible for large $\lambda$ and $\tan\beta \lsim 6$; subsequently this
region will be denoted as ``large $\lambda$'' (LLAM) region.
The region where contributions to $\Delta_\mathrm{NMSSM}$ from $H_{SM}' - S'$ mixing
are sizeable (up to $\sim 8$ GeV) is
characterised by a small value of $\lambda$ and large $\tan\beta$.
Subsequently we call this region the ``large mixing'' (LMIX) region. 

\begin{figure}[!h]
\centering
\includegraphics[scale=0.6]{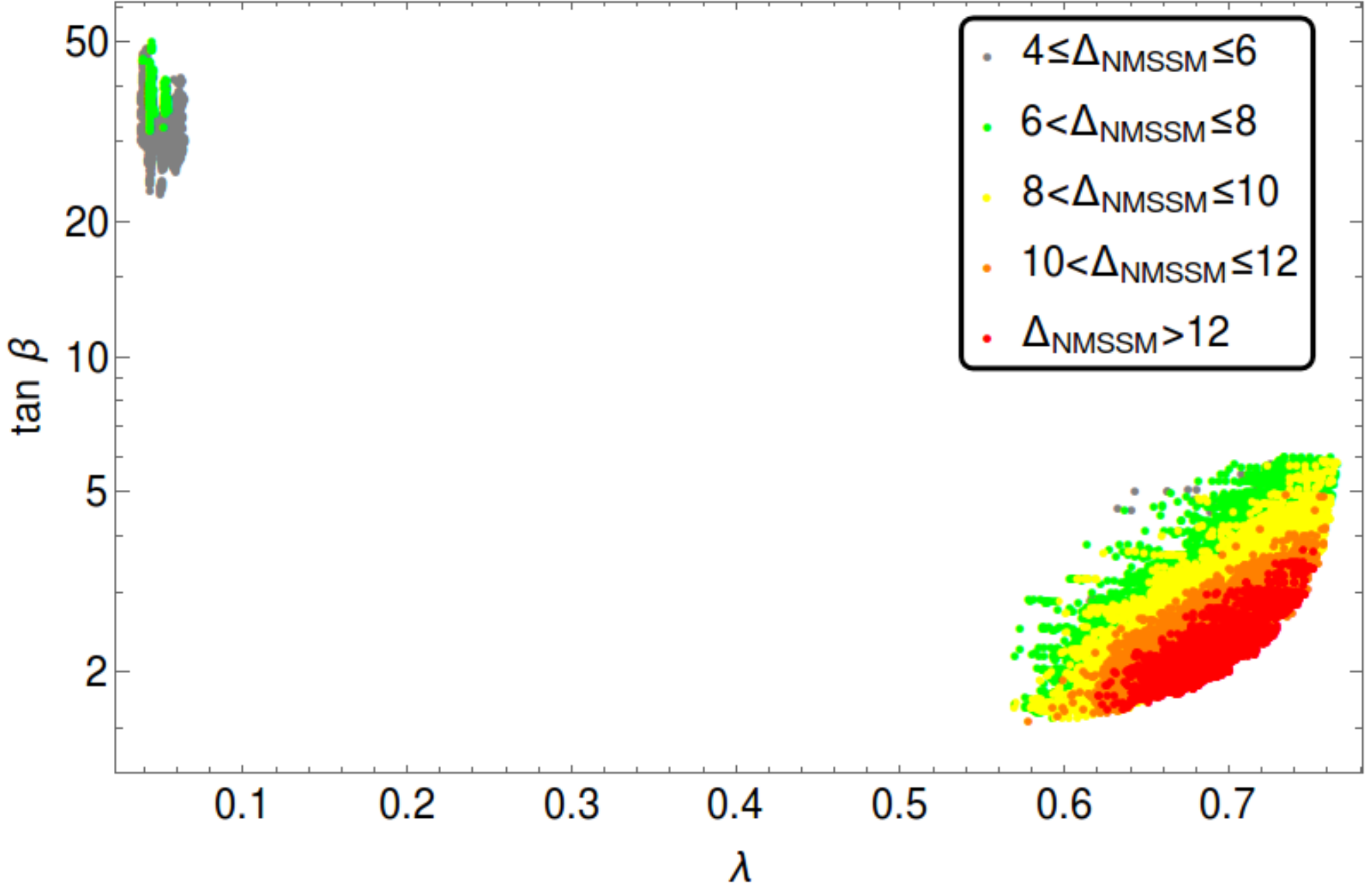}
\caption{$\lambda -\tan\beta$~plane showing the viable points and
$\Delta_{\text{NMSSM}}$ in
the form of a color code. The island in the upper-left corner corresponds to the
region where $\Delta_{\text{NMSSM}}$ originates from $H_{SM}' - S'$ mixing (LMIX),
whereas the island in
the large $\lambda$ regime (LLAM) corresponds to the region with large contributions to
$\Delta_{\text{NMSSM}}$ from the second term in (\ref{eq:m11}).}
\label{fig:lambdatanb}
\end{figure}

The viable points are shown in the $\lambda-\tan\beta$~plane in Fig.~\ref{fig:lambdatanb},
including the possible values of $\Delta_{\text{NMSSM}}$ in the form of a color code.
One can clearly distinguish the two ``islands'' of valid points in the
plane which can lead to a substantially different phenomenology, but both featuring a lower fine
tuning than in the MSSM. In the following subsections we show various observables
which can help to test these scenarios.
%
\subsection{Searches for $H_S$ in the diphoton final state}
As already stated above, the ATLAS and CMS collaborations have recently published results from
searches for additional BSM Higgs bosons with masses below 125~GeV  in the diphoton final state
\cite{Aad:2014ioa,CMS-HIG-14-037}, leading to upper bounds on corresponding signal rates.
First we have to verify whether these upper bounds lead to constraints on the parameter
space of the NMSSM considered above. To this end we have used the
public code \texttt{SusHi 1.5} \cite{Harlander:2012pb} to obtain the NNLO
gluon fusion production cross section for a SM-like Higgs boson, and multiplied it by the
reduced coupling of $H_S$ to gluons $\kappa^2_{gg}(H_S)$
given by the output of NMSSMTools. Finally the production cross section is
multiplied by the $BR(H_S\to \gamma\gamma)$ as given by NMSSMTools.

\begin{figure}[!ht]
\centering
\includegraphics[width=75mm]{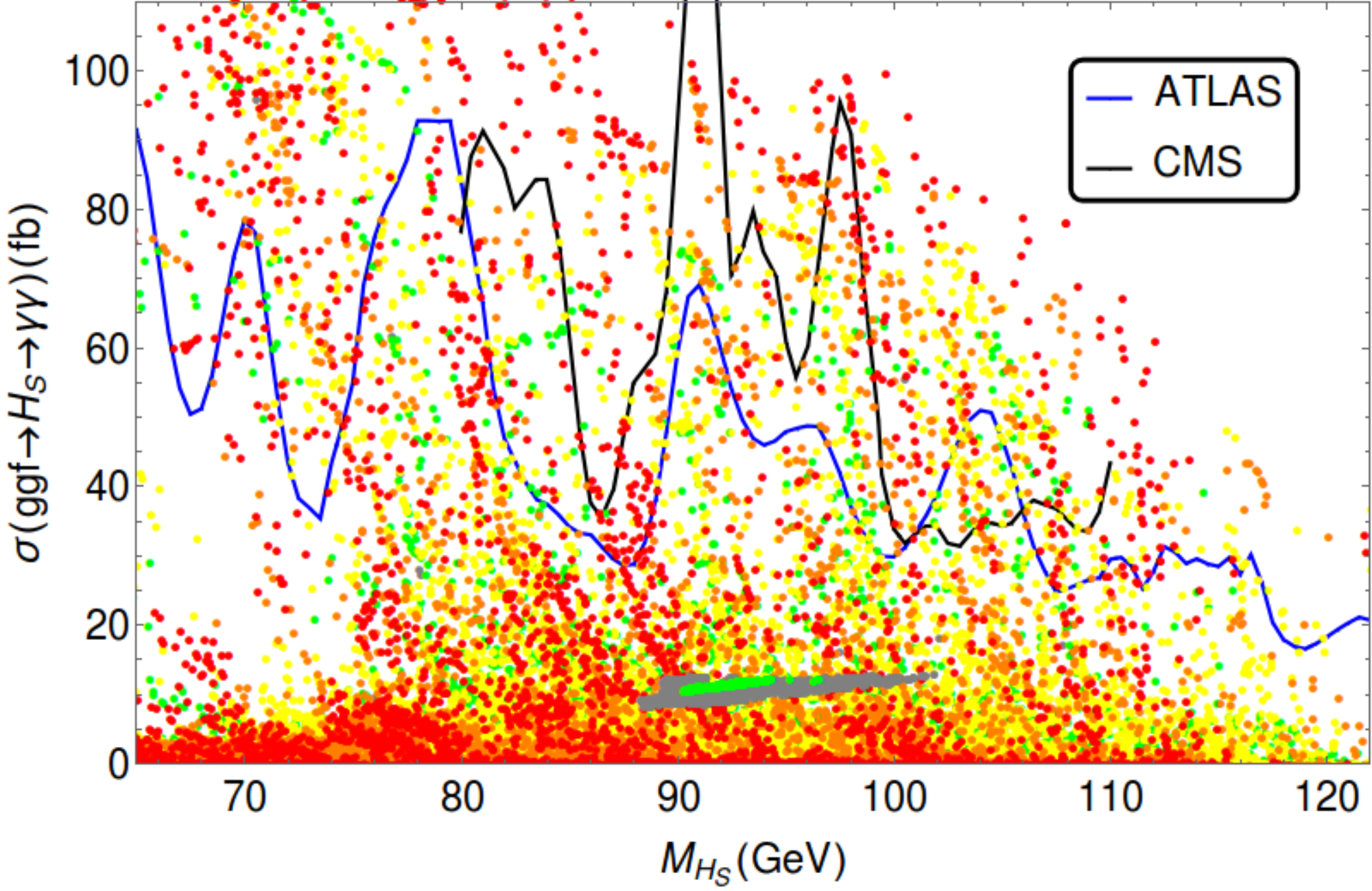}
\includegraphics[width=75mm]{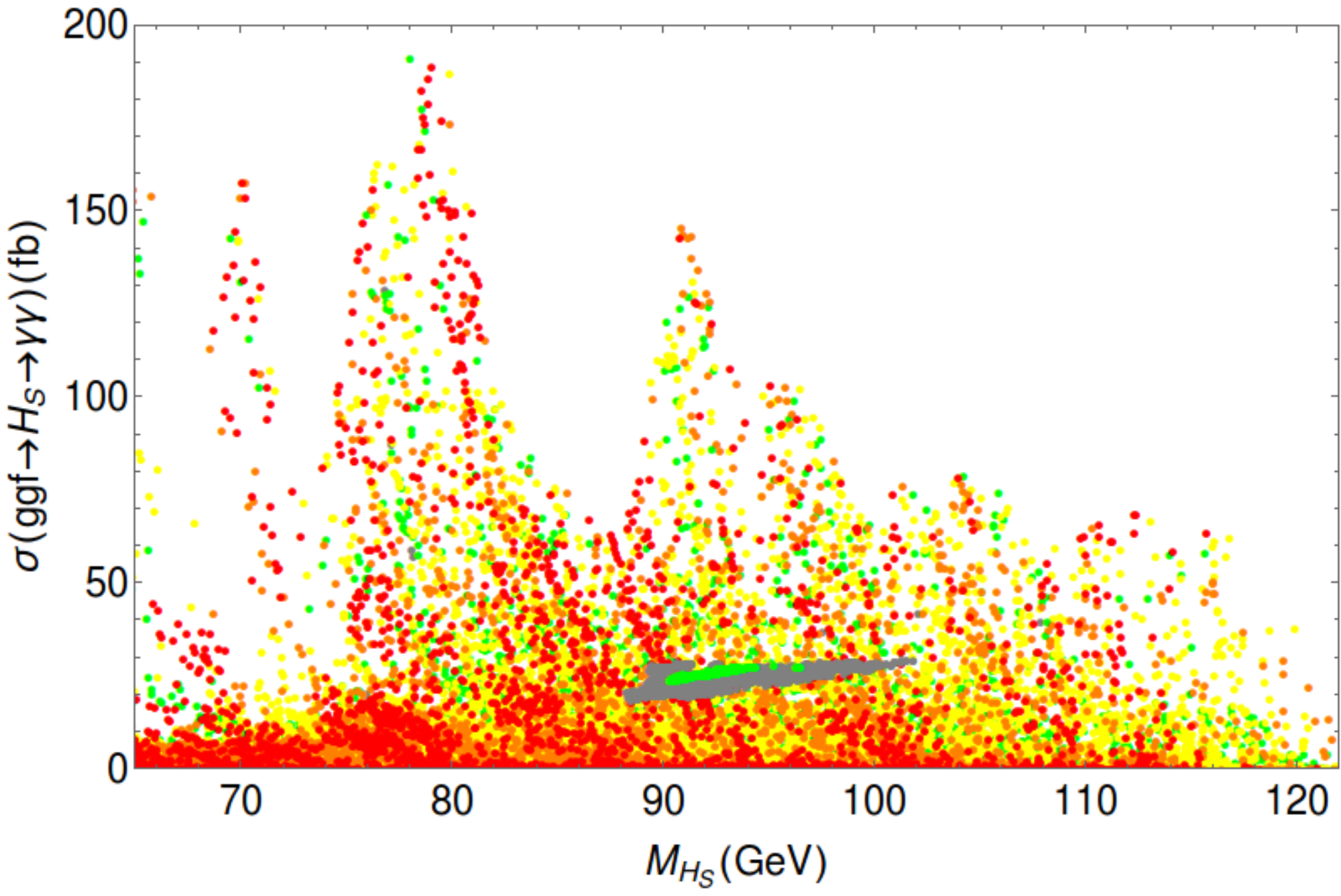}
\caption{Left: Possible signal rates (in femtobarns)
$\sigma({gg\rightarrow H_S \rightarrow \gamma\gamma})$
at a c.m. energy of $\sqrt{s}=8$ TeV, together with the ATLAS \cite{Aad:2014ioa}
and CMS \cite{CMS-HIG-14-037} limits from
direct searches. The grey-green island corresponds to the LMIX region, the rest to the
LLAM region. Right: Signal rates for the same
process at $\sqrt{s}=13$ TeV for the remaining points once the upper bounds from ATLAS and CMS
have been applied.}
\label{fig:gammagamma}
\end{figure}

On the left hand side of
Fig. \ref{fig:gammagamma} we show the resulting signal rates at $\sqrt{s}=8$ TeV c.m.
energy, together with the ATLAS \cite{Aad:2014ioa} and CMS \cite{CMS-HIG-14-037} limits from
direct searches as function of $M_{H_S}$. Here the LMIX region apprears as a grey-green island
within the much larger LLAM region.
On the right hand side of Fig. \ref{fig:gammagamma}
we show the resulting signal rates at
$\sqrt{s}=13$ TeV c.m. energy, once the constraints from ATLAS and CMS searches have
been applied.

We see in Figs.~\ref{fig:gammagamma} that in the grey-green LMIX region $M_{H_S}$ is confined
to the mass range 
\newline
$88\ \text{GeV} \lsim M_{H_S} \lsim 102\ \text{GeV}$, a consequence of
the parameter range \eqref{eqn:stopmassrange} and the corresponding lower limit
on $\Delta_{\text{NMSSM}}\gsim 4$~GeV. In order to obtain such values of $\Delta_{\text{NMSSM}}$
through $H_{SM}'-S'$ mixing, the mixing angle has to be relatively large leading to sizeable
couplings of $H_S$ to electroweak gauge bosons. These, in turn, are allowed by LEP only
in the corresponding mass range where, actually, a mild excess of events is
seen \cite{Schael:2006cr}.

The recent ATLAS and CMS searches have
not yet been sensitive to the possible signal rates $\sigma({gg\rightarrow H_S \rightarrow \gamma\gamma})$
in the LMIX region of the NMSSM, due to the absence of a possible enhancement of
the $BR(H_S\to \gamma\gamma)$
(see below). Fig.~\ref{fig:gammagamma} (right) indicates, on the other hand, that the LMIX region
could be completely tested once searches at $\sqrt{s}=13$~TeV c.m. energy become sensitive
to $\sigma({gg\rightarrow H_S \rightarrow \gamma\gamma}) \sim 20$~fb.

Within the LLAM (large $\lambda$) region both $M_{H_S}$ and 
$\sigma({gg\rightarrow H_S \rightarrow \gamma\gamma})$ can vary over much larger ranges and,
indeed, the ATLAS and CMS searches have started to test parts of the LLAM region
where this signal rate is particularly large. On the other hand this signal rate can
also be quite small in the LLAM region where $H_{SM}' - S'$ mixing is possible, but not
mandatory. This part of the LLAM region will be hard to test via searches for direct
$H_S$ production.

\begin{figure}[!hb]
\centering
\includegraphics[width=75mm]{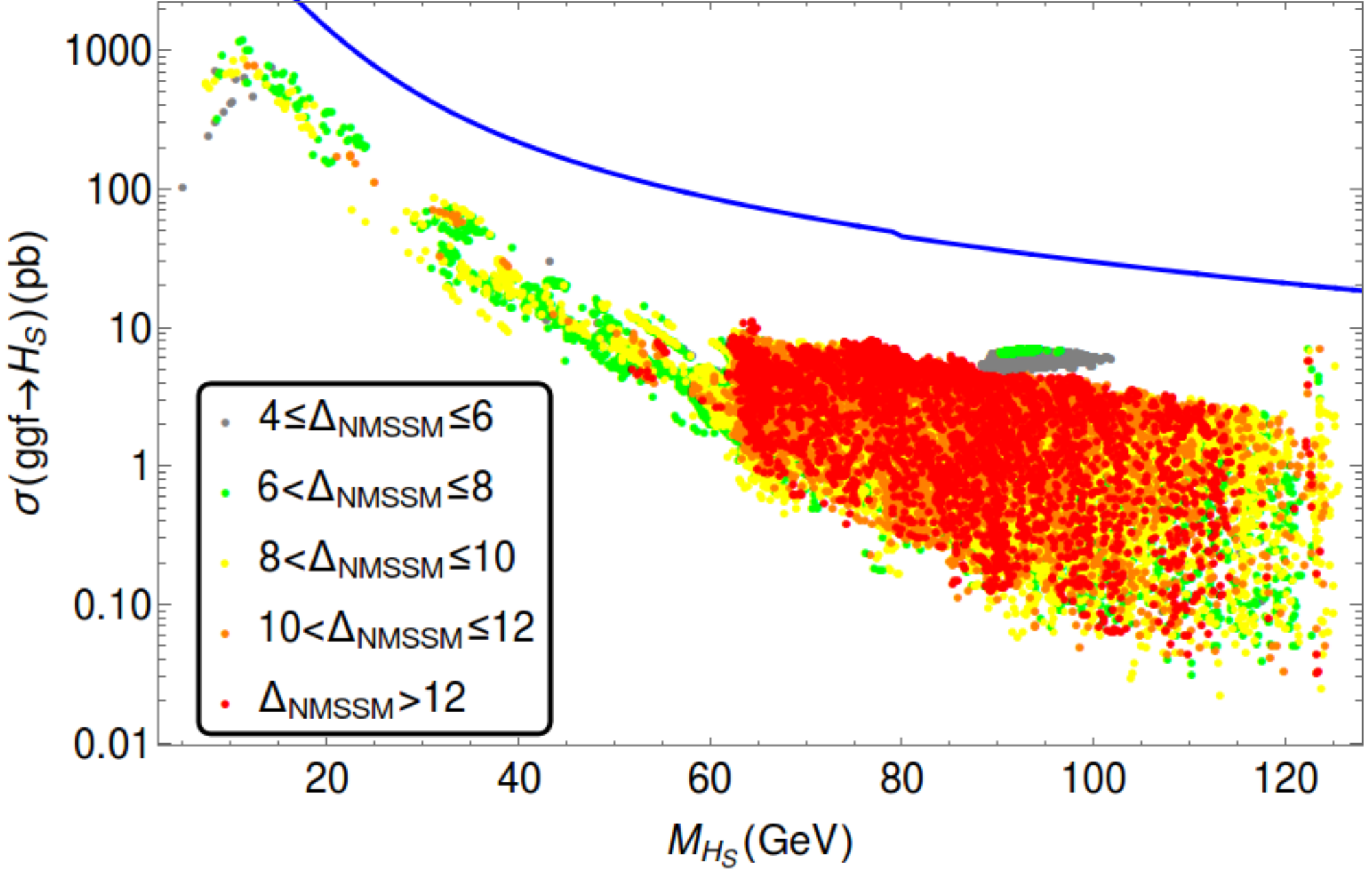}
\includegraphics[width=75mm]{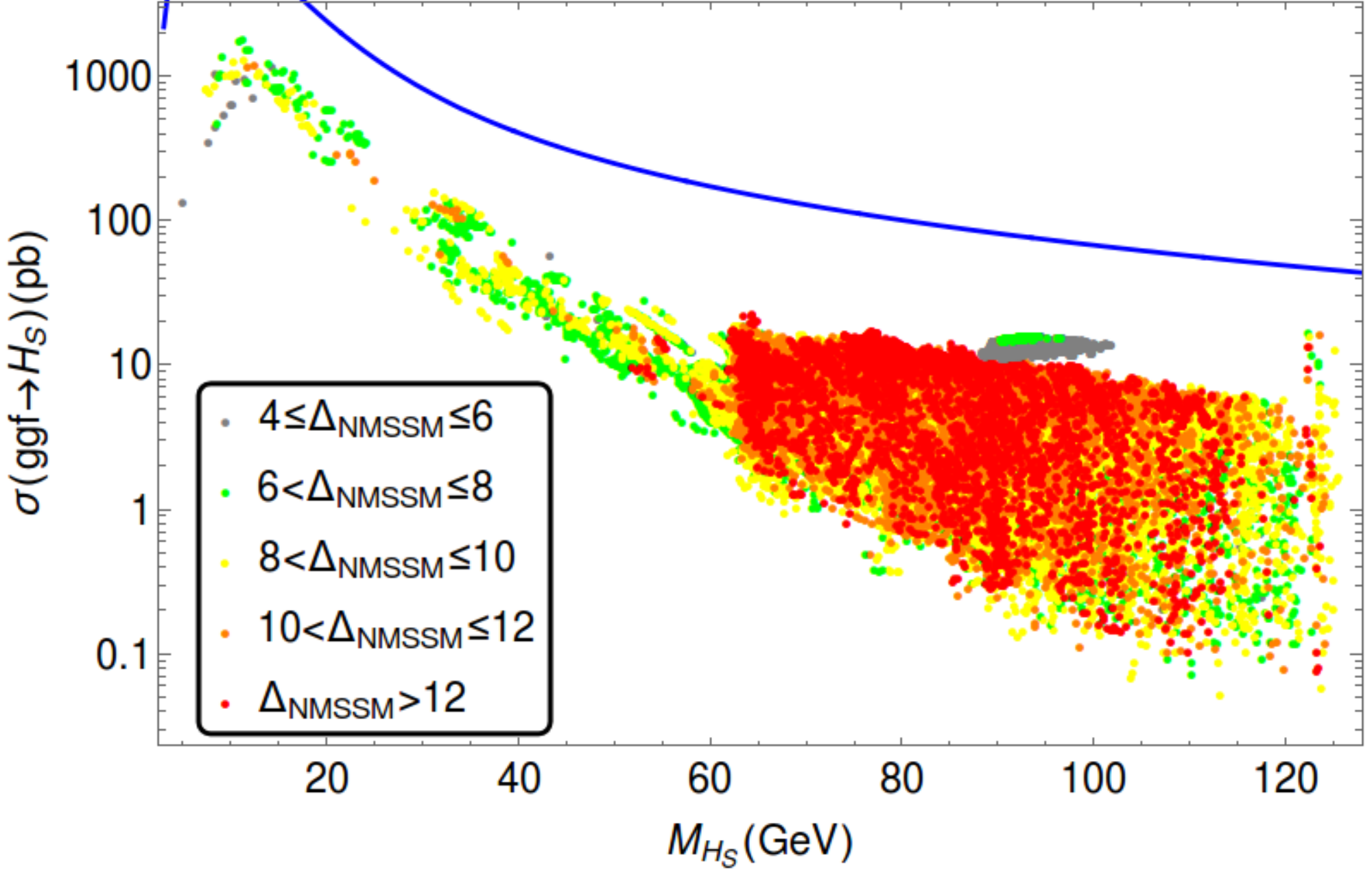}
\caption{Production cross section of $H_S$ at $\sqrt{s}=8$ TeV (left) and $\sqrt{s}=13$ TeV
(right) with the color code for $\Delta_{\text{NMSSM}}$. The blue line indicates the
corresponding ggF cross section for a SM Higgs boson of the same mass. The grey-green island
corresponds to the LMIX region.}
\label{fig:ggFhsproduction}
\end{figure}

\begin{figure}[!hb]
\centering
\includegraphics[width=75mm]{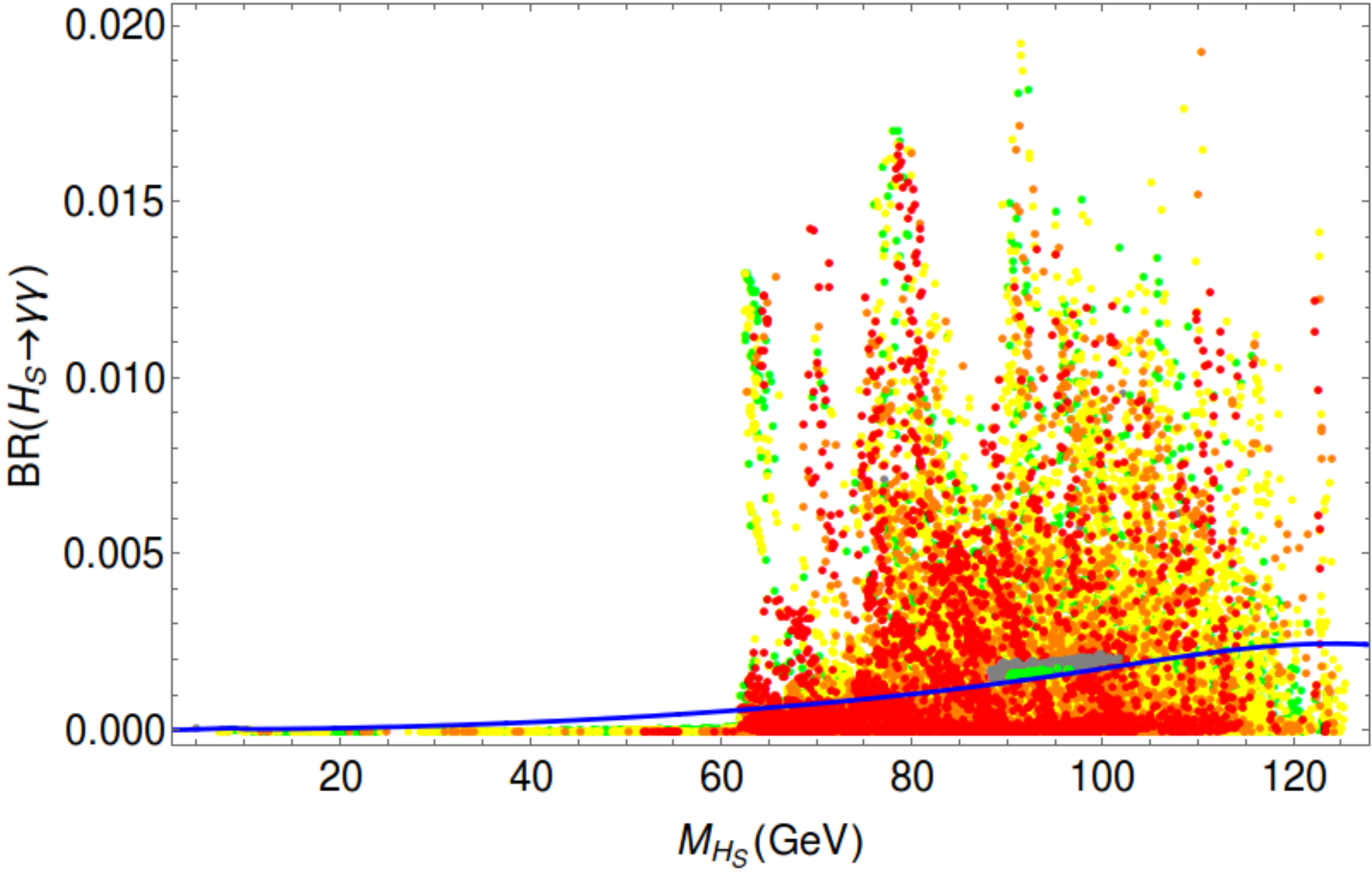}
\includegraphics[width=75mm]{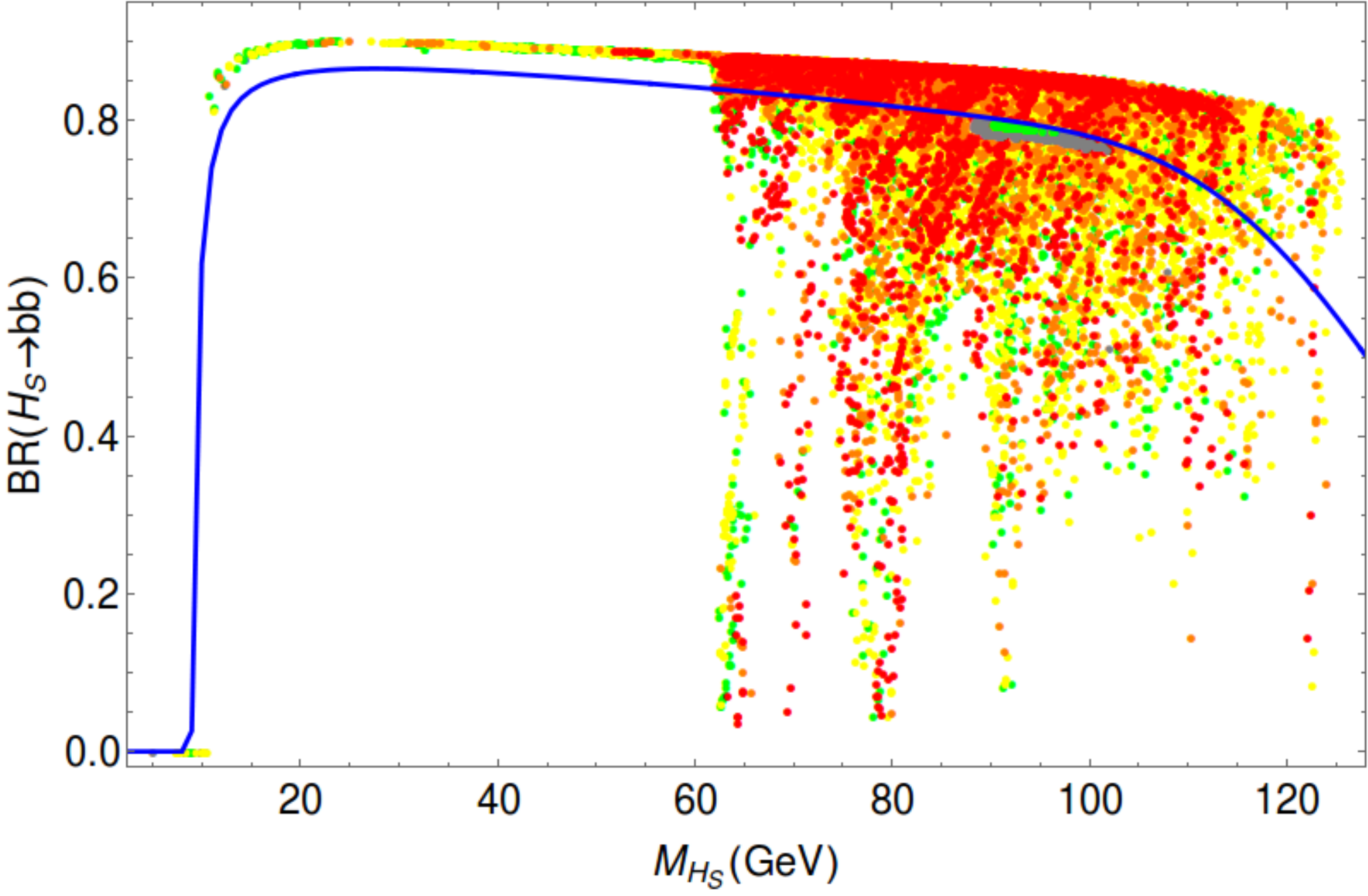}
\caption{Branching ratios of $H_S$ into photons (left) and $b\bar b$ (right) versus its mass.
The blue line indicates the
corresponding branching ratios for a SM Higgs boson of the same mass. The grey-green island
corresponds to the LMIX region, in which the branching ratios are very SM-like.}
\label{fig:hsBR}
\end{figure}

It is interesting to decompose $\sigma({gg\rightarrow H_S \rightarrow \gamma\gamma})$
into production cross sections and branching fractions, which allows to estimate signal rates
in other channels and to understand the origin of the varying signal rates in
Fig.~\ref{fig:gammagamma}. In Fig.~\ref{fig:ggFhsproduction} we show the production
cross section of $H_S$ at $\sqrt{s}=8$ TeV (left) and $\sqrt{s}=13$ TeV
(right) with the same color code for $\Delta_{\text{NMSSM}}$
as in Fig.~\ref{fig:lambdatanb}, omitting the points
excluded by ATLAS or CMS. We observe that, for
the allowed mass range $88\ \text{GeV} \lsim M_{H_S} \lsim 102\ \text{GeV}$,
$\sigma({gg\rightarrow H_S})$ is indeed larger in the LMIX region than in the
LLAM region, since the couplings of $H_S$ to fermions (here: to the top quark) are
relatively large.
However, the $BR(H_S\to \gamma\gamma)$ shown on the left hand side of
Fig.~\ref{fig:hsBR} clarify that these
can be (much!) larger for $H_S$ than for a SM-like Higgs (shown as blue line) only for
parts of the LLAM region, never within the LMIX region; only within the LLAM region a
suppression of the $BR(H_S\to b\bar{b})$ is possible (as shown on the right hand side of
Fig.~\ref{fig:hsBR}) which is required in order to enhance the $BR(H_S\to \gamma\gamma)$.

Finally both Figs.~\ref{fig:ggFhsproduction} and \ref{fig:hsBR} show that very few viable
points exist for $M_{H_S} < 60$~GeV (in the LLAM region only):
Such light states can be produced in decays
$H_{SM}\to H_S H_S$ and would reduce the
observed $H_{SM}$ signal rates into SM-like final states to inadmissible levels.
The $H_{SM} -H_S- H_S$ coupling can be small for large $\lambda$, however,
due to (rare) accidential cancellations among the various contributing terms.
(This mass range has not been shown in Fig.~\ref{fig:gammagamma} since the experiments
have not been sensitive to it.)

\subsection{Reduced Couplings of $H_{SM}$}

As stated above the LMIX (and LLAM) regions can have an impact on the
reduced couplings of $H_{SM}$, actually both due to $H_{SM}'-S'$ mixing
and $H_{SM}'-H'$ mixing induced by the final diagonalisation of the mass
matrix $\M'^2$ \eqref{2.10e}. The ATLAS and CMS measurements of the
reduced couplings of $H_{SM}$ at the first run of the LHC
have recently been combined in \cite{ATLAS-CMS}, and prospects for
future measurements have been published in
\cite{ATL-PHYS-PUB-2013-014}~(ATLAS) and \cite{CMS:2013xfa}~(CMS).

\begin{figure}[!b]
\centering
\includegraphics[scale=0.6]{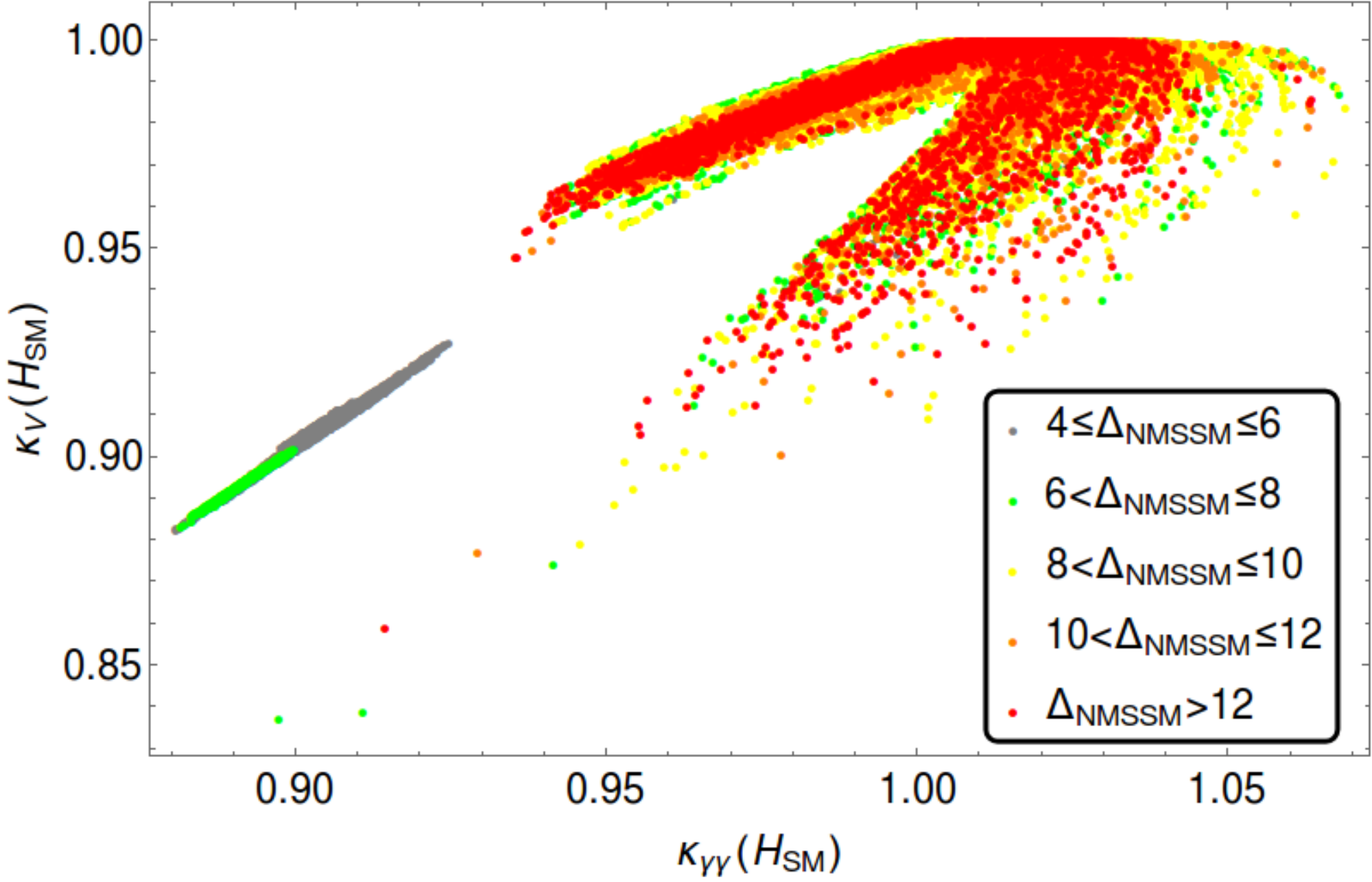}
\caption{Reduced couplings
$\kappa_V(H_{SM})$ and $\kappa_{\gamma\gamma}(H_{SM})$ for the viable points,
including a color code for $\Delta_{\text{NMSSM}}$.}
\label{fig:correlationsgggg}
\end{figure}

First we show in Fig.~\ref{fig:correlationsgggg} the reduced couplings
$\kappa_V(H_{SM})$ and $\kappa_{\gamma\gamma}(H_{SM})$ for the viable points.
The LMIX and LLAM regions can be distinguished
clearly in Fig.~\ref{fig:correlationsgggg}: As before the LMIX region corresponds
to the thin grey-green strip, the LLAM region to the remaining part
dominated by mostly red points (for which $12~\mathrm{GeV} < \Delta_{\text{NMSSM}} < 17$~GeV).

From the recent ATLAS-CMS combination in \cite{ATLAS-CMS} one finds for the
scenario relevant here (custodial
symmetry, i.e. $\kappa_Z(H_{SM})=\kappa_W(H_{SM})\equiv\kappa_V(H_{SM})\leq 1$)
that $\kappa_V(H_{SM})\gsim 0.83$ at the 95\%~CL level. The prospects
for the measurements of $\kappa_V(H_{SM})$ at the run~II of the LHC  in
\cite{ATL-PHYS-PUB-2013-014}~(ATLAS) and \cite{CMS:2013xfa}~(CMS) depend on
uncertainty scenarios and, of course, on the integrated luminosity. For
300~fb$^{-1}$ one expects uncertainties of about 5\% at the $1\sigma$ level,
i.e. the possibility to set a lower bound on $\kappa_V(H_{SM})$ of
$\sim 0.9$ at the 95\%~CL level. Such a bound can test the green 
$\Delta_{\text{NMSSM}} > 6$~GeV region of the LMIX scenario, but reduced
uncertainties of about 7\%
at the 95\%~CL level at 3000~fb$^{-1}$ integrated luminosity could test the
LMIX scenario completely. Again, the LLAM scenario can be tested only
partially by measurements of $\kappa(H_{SM})$.
The prospects for constraining (or detecting) the LMIX/LLAM scenarios
via measurements of $\kappa_{\gamma\gamma}(H_{SM})$ are similar, but
somewhat less promising due to the larger foreseen uncertainties at
both 300~fb$^{-1}$ and 3000~fb$^{-1}$ integrated luminosity
\cite{ATL-PHYS-PUB-2013-014,CMS:2013xfa}.

\begin{figure}[ht!]
\centering
\includegraphics[width=75mm]{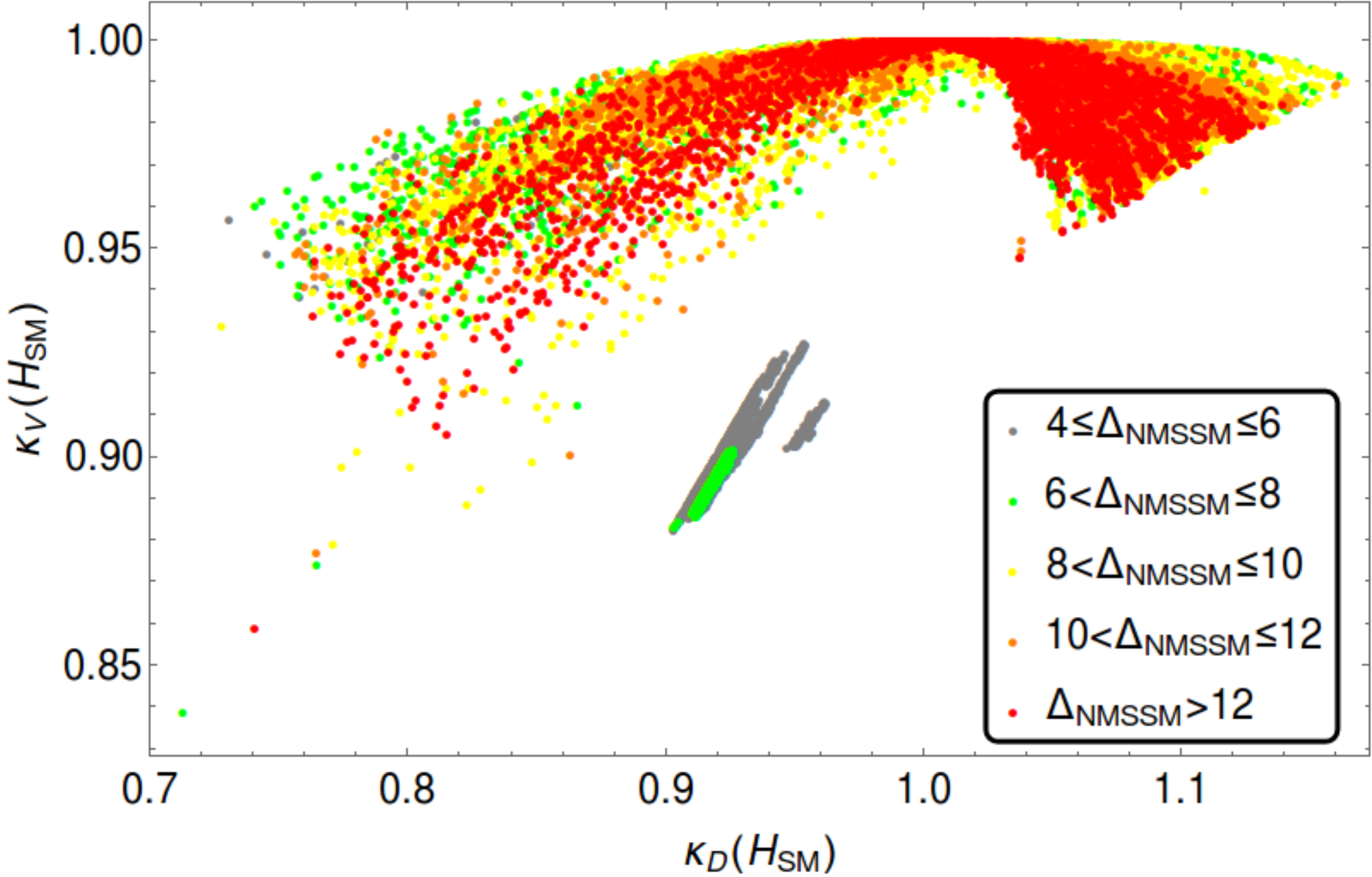}
\includegraphics[width=75mm]{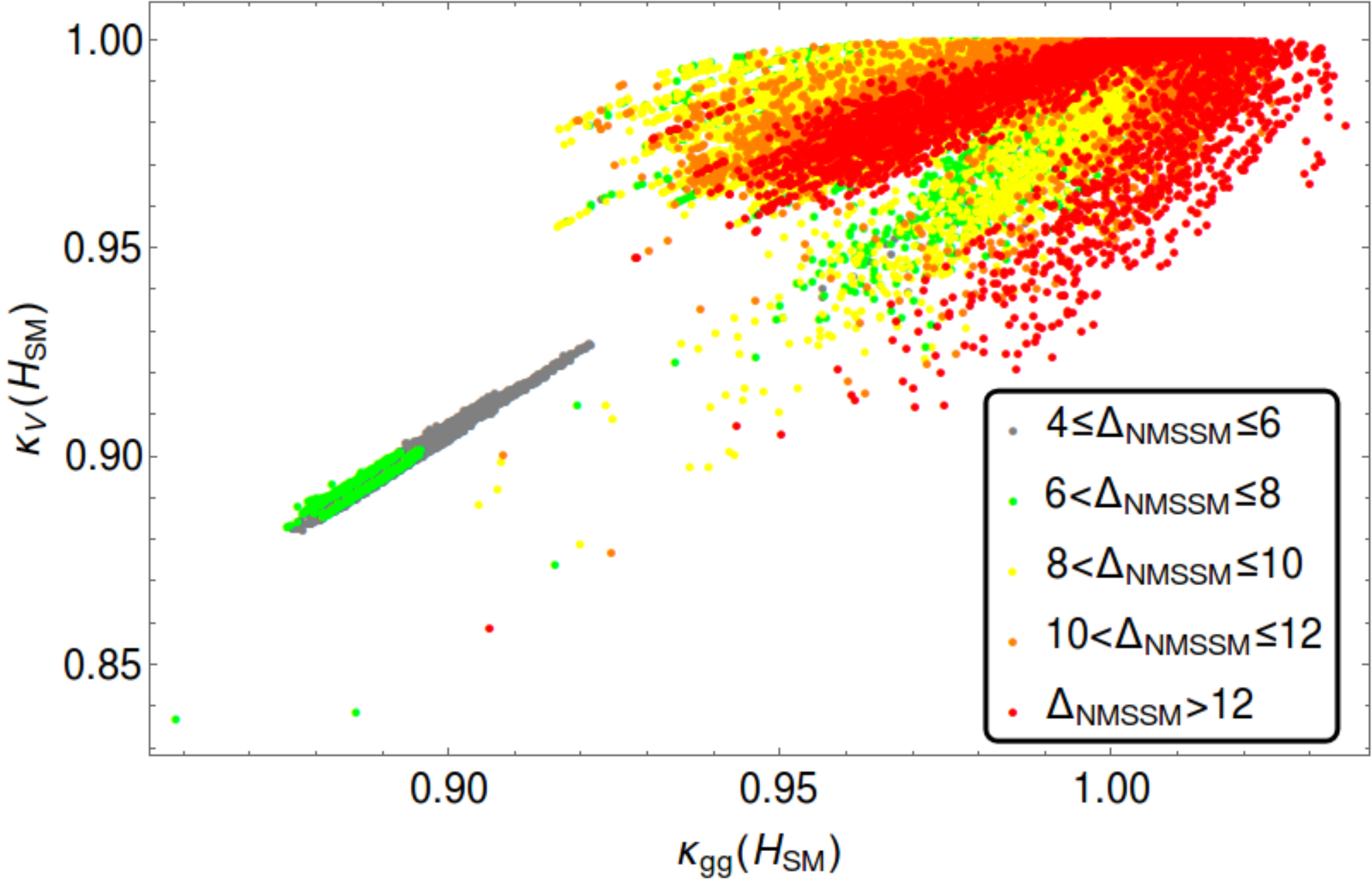}
\caption{Correlations of $\kappa_V(H_{SM})$ with
the reduced couplings of $H_{SM}$ to down-type fermions (left) and gluons (right).
In both plots the two regions LLAM and LMIX are clearly separated.}
\label{fig:correlationsfermions}
\end{figure}

Apart by future measurements of individual values of reduced couplings
of $H_{SM}$, informations or constraints on scenarios predicting
deviations from the SM can be obtained by considering correlations
among reduced couplings. To this end we show in Figs.~\ref{fig:correlationsfermions}
the correlations of $\kappa_V(H_{SM})$ with
the reduced couplings of $H_{SM}$ to down-type fermions ($\kappa_D(H_{SM})$)
and gluons ($\kappa_{gg}(H_{SM})$).

Like in Fig.~\ref{fig:correlationsgggg} these correlations
are very pronounced in the LMIX scenario, but in the LLAM scenario
a wide range of $\kappa_{D}(H_{SM})$ is possible: a reduction of the
coupling of $H_{SM}$ to down-type fermions originates from negative
contributions to this coupling from $H_{SM}'-H'$~mixing. As for
$H_S$, a corresponding reduction of the $BR(H_{SM} \to b\bar{b})$
can lead to an enhanced $BR(H_{SM} \to \gamma\gamma)$ as observed
in Fig.~\ref{fig:correlationsgggg}. However, positive contributions
to the coupling of $H_{SM}$ to down-type fermions are possible as well,
with opposite consequences. The two regions $\kappa_{D}(H_{SM})>1$
and $\kappa_{D}(H_{SM})<1$ explain the origin of the two ``branches''
of $\kappa_V(H_{SM})$ visible in Fig.~\ref{fig:correlationsgggg} as
well on the right hand side of Fig.~\ref{fig:correlationsfermions}.
Unfortunately, the couplings of $H_{SM}$ can also be very SM-like,
like in the alignment limit studied recently in \cite{Carena:2015moc}.

\begin{figure}[ht!]
\centering
\includegraphics[width=75mm]{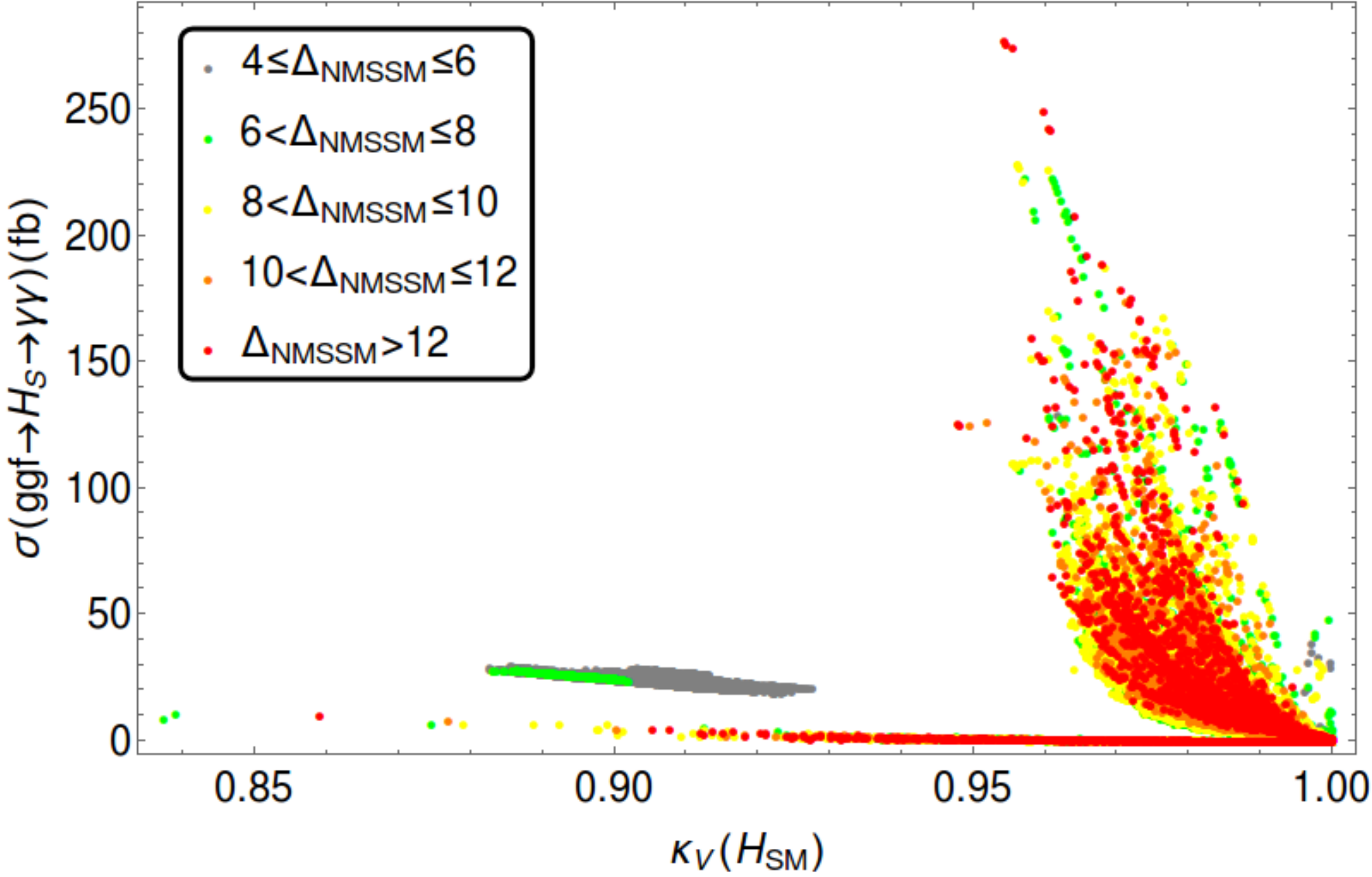}
\includegraphics[width=75mm]{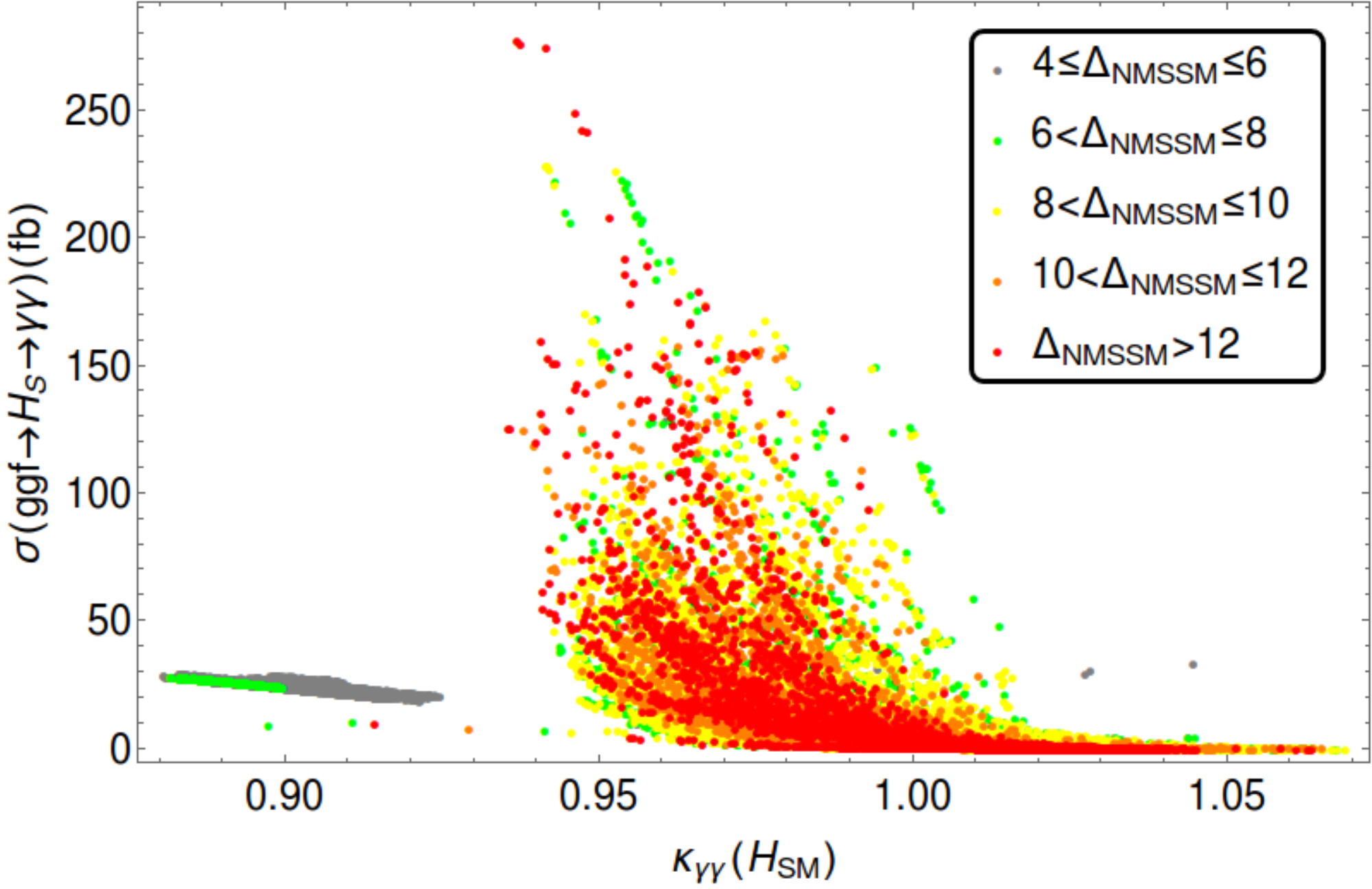}
\caption{Left: Correlations among the diphoton signal rate of $H_S$ and
$\kappa_{V}(H_{SM})$. Right: Correlations among the diphoton signal rate of $H_S$
and $\kappa_{\gamma\gamma}(H_{SM})$.}
\label{fig:correlationsignals}
\end{figure}

Next we turn to correlations between the reduced couplings of $H_{SM}$ and
the signal rates $\sigma({gg\rightarrow H_S \rightarrow \gamma\gamma})$
discussed in the previous subsection. In Figs.~\ref{fig:correlationsignals}
we show $\sigma({gg\rightarrow H_S \rightarrow \gamma\gamma})$ against
$\kappa_{V}(H_{SM})$ (left) and $\sigma({gg\rightarrow H_S \rightarrow \gamma\gamma})$
against $\kappa_{\gamma\gamma}(H_{SM})$ (right). These figures allow
to verify the possible complementarity of  measurements of
$\sigma({gg\rightarrow H_S \rightarrow \gamma\gamma})$ and the reduced
couplings of $H_{SM}$: In order to test the LMIX region (the grey-green island
on the left hand side), the necessary limits on 
$\sigma({gg\rightarrow H_S \rightarrow \gamma\gamma})$ and/or $\kappa_{V}(H_{SM})$
can now be deduced together. The LLAM region can become
visible either by an enhanced $\sigma({gg\rightarrow H_S \rightarrow \gamma\gamma})$
or a reduced $\kappa_{V}(H_{SM})$, but not both. Unfortunately,
a low signal rate $\sigma({gg\rightarrow H_S \rightarrow \gamma\gamma})$ as well as
$\kappa_{V}(H_{SM}) \sim 1$ are possible simultaneously.
From the right hand side of Figs.~\ref{fig:correlationsignals} we see that
enhanced signal rates 
$\sigma({gg\rightarrow H_S \rightarrow \gamma\gamma}) \gsim 50$~fb and enhanced reduced
couplings $\kappa_{\gamma\gamma}(H_{SM})$ are incompatible in the LLAM
region.

\subsection{$H_S$ production via decays of heavy states $H/A$}

\begin{figure}[hb!]
\centering
\includegraphics[scale=0.7]{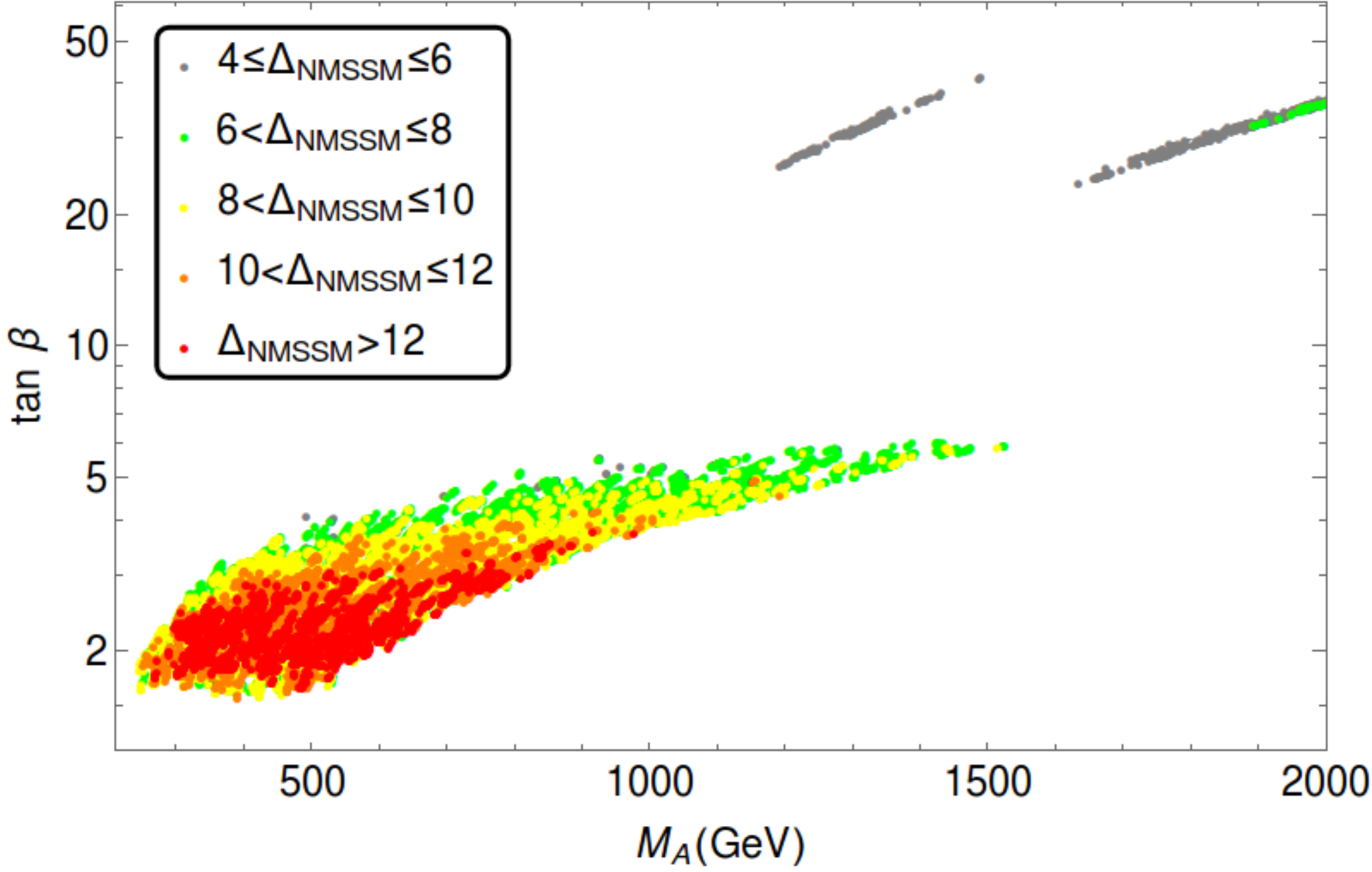}
\caption{Viable points in the $\tan\beta - M_A$ plane.}
\label{fig:mAtanb}
\end{figure}

Another way to produce a light $H_S$ is through the decays of
heavy (MSSM-like) states $H/A$. First we have to find out which
masses of $H/A$ are possible in the LMIX/LLAM regions of the NMSSM
considered here. In Fig.~\ref{fig:mAtanb} we show the regions of
viable points in the $\tan\beta - M_A$ plane, which helps to
clarify that these points are not ruled out by searches for MSSM-like
$H/A$ in the $\tau^+\tau^-$ final state
(from here onwards, $M_A$ denotes the physical mass of the MSSM-like
CP-odd state $A$): The LMIX region with large $\tan\beta$ features
very heavy $H/A$ states, to which searches at the run~I have not been
sensitive (and which will be hard to search for at the run~II).
The LLAM region is characterized by lower $\tan\beta$ such that the
associate production of $H/A$ states with $b$~quarks is not very
enhanced; instead, their production via gluon fusion becomes
feasable in principle \cite{Djouadi:2013vqa}. The part of the LLAM region
where $M_A \gsim 500$~GeV and $\tan\beta \gsim 3$ corresponds, however,
to the difficult region where the reduced couplings of $H_{SM}$ are
very SM-like and $H_S$ has a low signal rate in the $\gamma\gamma$
channel; in this region also the search for
the MSSM-like states $H/A$ seems difficult \cite{Djouadi:2015jea}.

\begin{figure}[ht!]
\centering
\includegraphics[width=75mm]{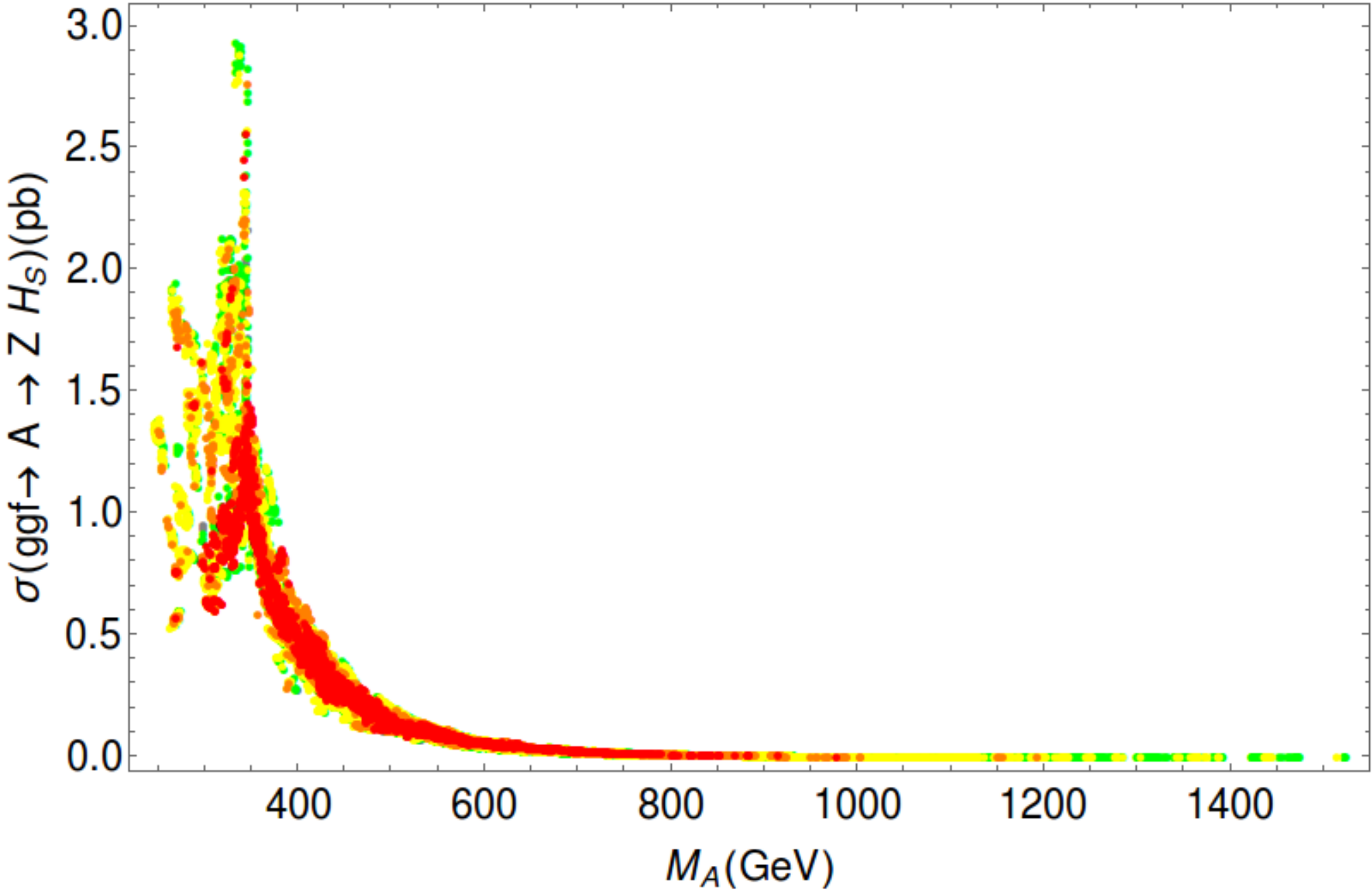}
\includegraphics[width=75mm]{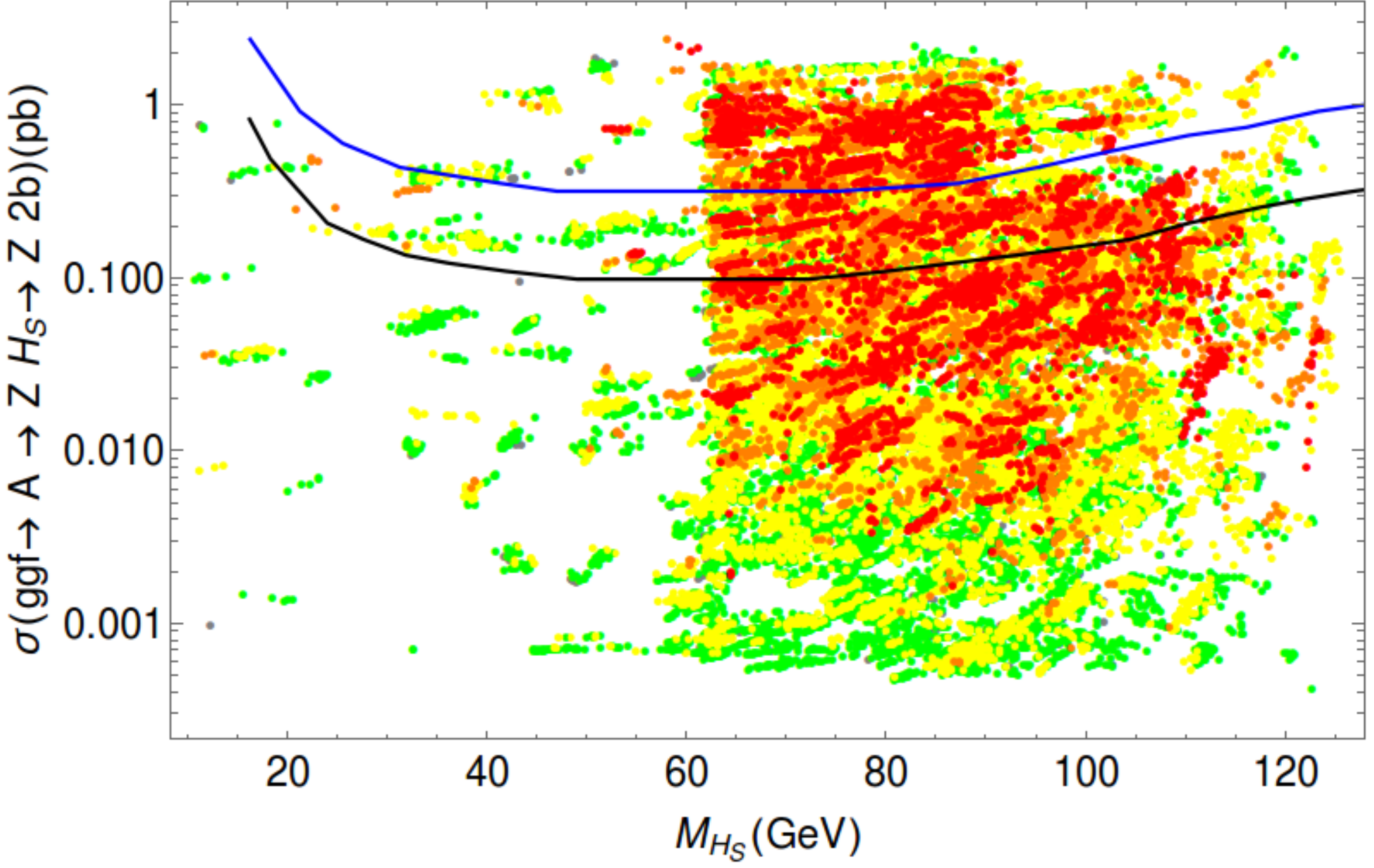}
\caption{Left: Signal cross section
$\sigma(ggF\to A \to Z+H_S)$ as function of $M_A$
for a c.m. energy of $\sqrt{s}=13$~TeV. Right: Signal cross section
$\sigma(ggF\to A \to Z+b+\bar{b})$ as function of $M_{H_S}$,
compared to the expected sensitivities
for a integrated luminosities of
$300~\text{fb}^{-1}$ (blue) and $3000~\text{fb}^{-1}$ (black)
(from \cite{Bomark:2014gya}).}
\label{fig:decayAZh}
\end{figure}

Promising decays of $H/A$ into $H_S$ are $A \to Z + H_S$ and $H \to H_{SM}+H_S$.
Since the kinematics of $A \to Z + H_S$ is very similar to the one of
$H \to Z + A_S$ investigated in \cite{Bomark:2014gya}, the studies
of the $Z \to l^+l^-$ ($l\equiv e,\mu$) and $A_S \to b\bar{b}$ final states
in \cite{Bomark:2014gya} can be employed, including their sensitivity
curves as function of $M_{A_S}$ (now interpreted as $M_{H_S}$). First
we show what signal cross sections can be
expected as function of $M_A$. The signal cross section
$\sigma(ggF\to A \to Z+H_S)$ is shown on the left hand side
of Fig.~\ref{fig:decayAZh} as function of $M_A$; clearly, visible signal rates can
only be expected for $M_A \lsim 400$~GeV within the LLAM region. On the
right hand side of Fig.~\ref{fig:decayAZh} the range of signal cross sections
$\sigma(ggF\to A \to Z+b+\bar{b})$ is shown as function of $M_{H_S}$,
and compared to the expected sensitivities at the run~II of the LHC
for integrated luminosities of
$300~\text{fb}^{-1}$ (blue) and $3000~\text{fb}^{-1}$ (black)
(from \cite{Bomark:2014gya}). Hence, detectable signal rates in this
channel are indeed possible in the LLAM region of the NMSSM without,
however, covering it completely.

\begin{figure}[hb!]
\centering
\includegraphics[width=75mm]{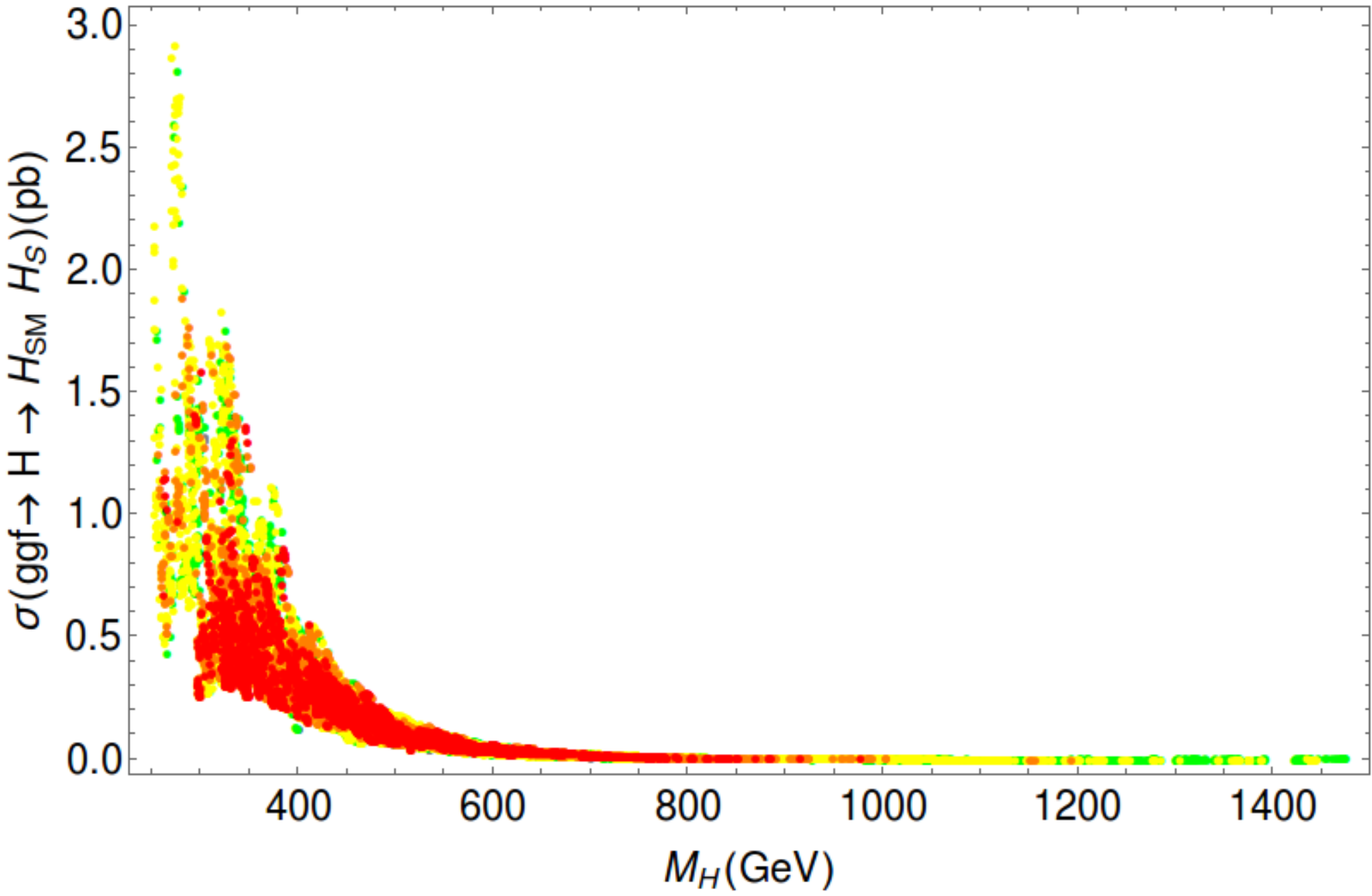}
\includegraphics[width=75mm]{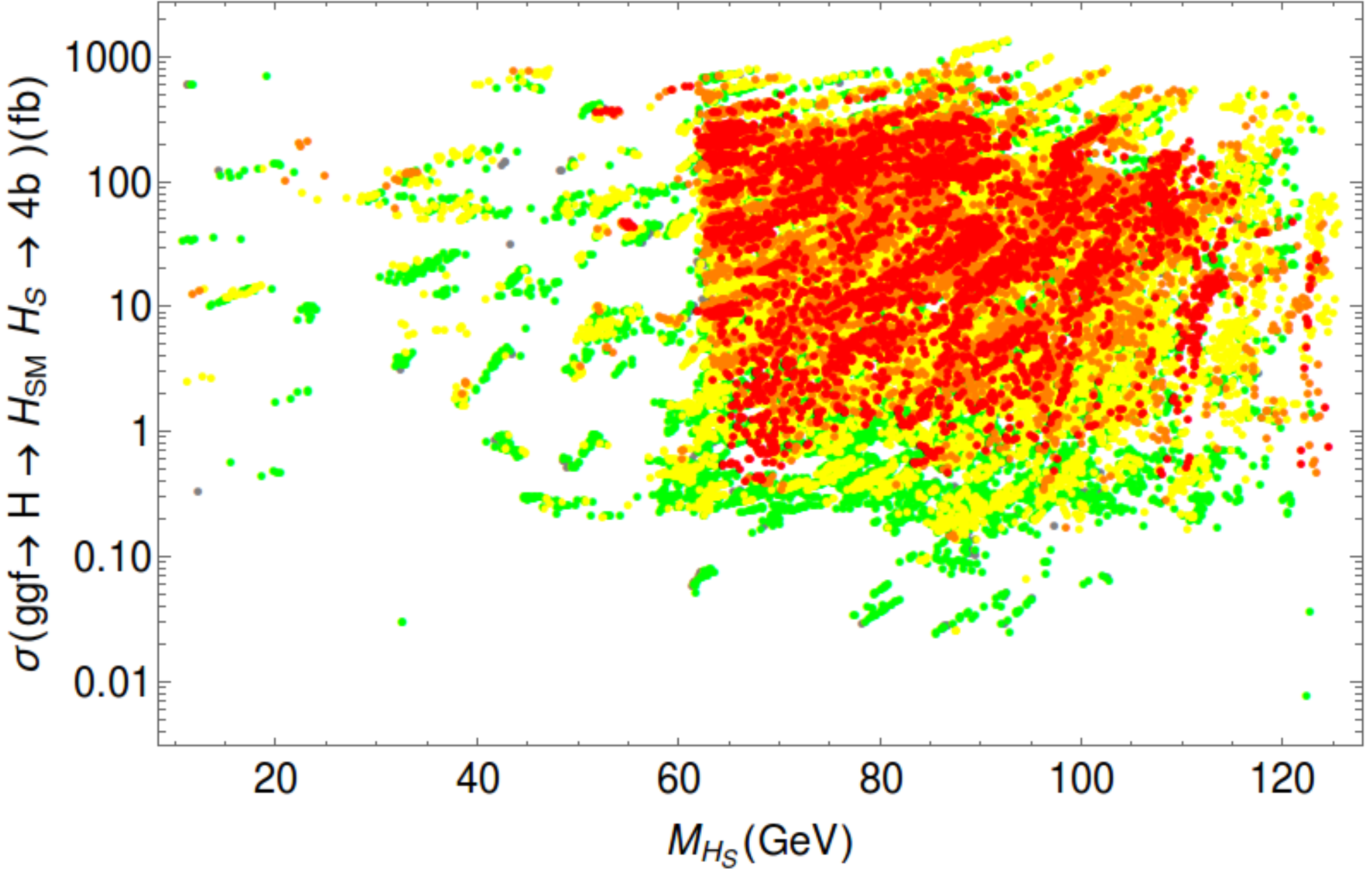}
\caption{Total ggF production cross section for $H\rightarrow H_{SM}+H_S$ at a c.m. energy
$\sqrt{s}=13$ TeV (left), and the signal cross section
into $b\bar b b \bar b$ versus the mass of $H_S$ (right).}
\label{fig:decaysH}
\end{figure}

The process $H \to H_{SM}+H_S$ can, in principle, be searched for
in various final states as $4b$, $2b2\tau$ and $4\tau$; one is handicapped,
however, by the a priori unknown mass of $H_S$.
In Fig.~\ref{fig:decaysH} we show the cross section
$\sigma(ggF\to H \to H_{SM}+H_S)$ as function of $M_H$
for a c.m. energy of $\sqrt{s}=13$~TeV on the left, and the (dominant)
signal cross section $\sigma(ggF\to H \to H_{SM}+H_S\to 4b)$ as function
of $M_{H_S}$ on the right. Search strategies including background studies
can possibly be persued along the lines proposed in \cite{Bomark:2014gya}
for searches for a light NMSSM pseudoscalar $A_S$. In the region of the
NMSSM parameter space considered here $A_S$ is, however, not particularly light;
we found that, in the (wider) LLAM region, $M_{A_S}$ varies from
$\sim 80$~to~$\sim 300$~GeV, but from $\sim 60$~to~$\sim 180$~GeV in the
(narrower) LMIX region. Search strategies including background studies
for searches for $H_S/A_S$ in Higgs-to-Higgs decays are
beyond the scope of the present paper and merit future studies.

\section{Conclusions}

We have studied a region in the NMSSM parameter space in which the mass
of the SM-like Higgs boson is uplifted by $\sim 4-17$~GeV, allowing for both
stop masses and $|A_t| \leq 1$~TeV alleviating the little fine tuning
problem of the MSSM. 
This region features a lighter mostly singlet-like Higgs state $H_S$
with a mass in the $60 - 125$~GeV range if the uplift is due to
singlet-doublet mixing (the LMIX region). 
Confining ourselves to values of $\lambda \lsim 0.75$,
this mass range of $H_S$ is
also natural in the LLAM region where the uplift originates from the
additional quartic term $\sim \lambda^2$ in the potential of the
SM-like Higgs boson.

The aim of the paper is the study of possible direct or indirect searches for
a light $H_S$
at the run~II of the LHC. Three possibilities have been considered:\\
a) Direct production of $H_S$ in gluon fusion, with $H_S$ decaying into
diphotons. Corresponding searches have been conducted recently by
ATLAS and CMS (the results of which have been taken into account), and
are the most promising also for the run~II of the LHC.\\
b) Modified reduced couplings of the SM-like Higgs state $H_{SM}$
through singlet-doublet mixing (both in the LMIX and the LLAM regions).\\
c) Production of $H_S$ in decays of heavier $H/A$ states, where we
confined ourselves to the most promising $A \to Z + H_S$
channel.

We found that the LMIX region can be tested if searches for BSM Higgs
bosons in the mass range $88 - 102$~GeV
become sensitive to signal cross sections
$\sigma({gg\rightarrow H_S \rightarrow \gamma\gamma}) \sim 20$~fb.
Alternatively, the LMIX region can be tested if measurements of the
reduced coupling $\kappa_V(H_{SM})$ of the SM Higgs boson to
electroweak gauge boson exclude (or confirm) the region
$\kappa_V(H_{SM}) \lsim 0.93$. Since the $H/A$ states are always
quite heavy in the LMIX region (with masses well above 1~TeV), $H_S$
detection via $H/A$ seems impossible in the near future, and tests
of the LMIX region have to rely on one of the two
measurements above, which seems feasable if the projected sensitivities
can be reached.

On the other hand it is difficult to test the entire LLAM region
even if $H_S$ is light (with a mass below 125~GeV), the range
considered here: both the signal cross section
$\sigma({gg\rightarrow H_S \rightarrow \gamma\gamma})$ and the
deviation of the reduced couplings of $H_{SM}$ from one can simultaneously
be very small. However, in other parts of the LLAM region
both the signal cross section
$\sigma({gg\rightarrow H_S \rightarrow \gamma\gamma})$ and the
deviation of the reduced couplings of $H_{SM}$ from one can be
much larger than in the LMIX region; these parts of the LLAM region
will be the first ones to be tested. In a part of the ``difficult''
LLAM region, but for which the $H/A$ states are not too heavy (with masses
$\lsim 400$~GeV), the detection of $H_S$ at least via
$ggF\to A \to Z+H_S$ is possible. Studies on the possible detection
of $H_S$ via other $H/A$ decay channels (including larger masses
of $H_S$) are planned.

\vfill

\section*{Acknowledgements}

U.E. and M.R. acknowledge support from the European Union Initial
Training Network Higgs\-Tools (PITN-GA-2012-316704)
and the D\'efi InPhyNiTi project N2P2M-SF.
U.E. acknowledges support from European Union Initial
Training Network INVISIBLES (PITN-GA-2011-289442),
the ERC advanced grant Higgs@LHC, and from
the grant H2020-MSCA-RISE-2014 No. 645722 (NonMinimalHiggs).

\end{document}